\documentclass[aps,prd,showpacs,twocolumn,superscriptaddress]{revtex4}
\usepackage{graphicx}

\def\lsim{\raise0.3ex\hbox{$<$\kern-0.75em\raise-1.1ex\hbox{$\sim$}}}
\def\gsim{\raise0.3ex\hbox{$>$\kern-0.75em\raise-1.1ex\hbox{$\sim$}}}

\begin{document}


\title{Charmonium spectral functions with the variational method \\ in zero and finite temperature lattice QCD}

\author{H.~Ohno}
\affiliation{Graduate School of Pure and Applied Sciences, University of Tsukuba,
	Tsukuba, Ibaraki 305-8571, Japan}
\author{S.~Aoki}
\affiliation{Graduate School of Pure and Applied Sciences, University of Tsukuba,
	Tsukuba, Ibaraki 305-8571, Japan}
\affiliation{Center for Computational Sciences, University of Tsukuba,
	Tsukuba, Ibaraki 305-8577, Japan}	
\author{S.~Ejiri}
\affiliation{Graduate School of Science and Technology, Niigata University,
	Niigata 950-2181, Japan}
\author{K.~Kanaya}
\affiliation{Graduate School of Pure and Applied Sciences, University of Tsukuba,
	Tsukuba, Ibaraki 305-8571, Japan}
\author{Y.~Maezawa}
\affiliation{Physics Department, Brookhaven National Laboratory,
	Upton, New York 11973, USA}
\author{H.~Saito}
\affiliation{Graduate School of Pure and Applied Sciences, University of Tsukuba,
	Tsukuba, Ibaraki 305-8571, Japan}
\author{T.~Umeda}
\affiliation{Graduate School of Education, Hiroshima University,
	Hiroshima 739-8524, Japan}

\collaboration{WHOT-QCD Collaboration}

\begin{abstract}
We propose a method to evaluate spectral functions on the lattice based on a variational method.
On a lattice with a finite spatial extent, spectral functions consist of discrete spectra only. 
Adopting a variational method, we calculate the locations and the heights of spectral functions at low-lying discrete spectra. 
We first test the method in the case of analytically solvable free Wilson quarks at zero and finite temperatures
and confirm that the method well reproduces the analytic results for low-lying spectra.
We find that we can systematically improve the results by increasing the number of trial states.
We then apply the method to calculate the charmonium spectral functions for S and P-wave states at zero-temperature in quenched QCD 
and compare the results with those obtained using the conventional maximum entropy method  (MEM). 
The results for the ground state are consistent with the location and the area of the first peak in spectral functions from the MEM, while
the variational method leads to a mass which is closer to the experimental value for the first excited state.
We also investigate the temperature dependence of the spectral functions for S-wave states below and above $T_c$.
We obtain no clear evidences for dissociation of $J/\psi$ and $\eta_c$ up to 1.4$T_c$.
\end{abstract}


\maketitle
\section{\label{sec:intro}Introduction}
Quark-gluon-plasma (QGP) is expected to be formed at sufficiently high temperatures and densities 
and considered to play an important role in the early universe and in the core of neutron stars.
In heavy-ion collision experiments, the QGP created is expected to cause a suppression of $J/\psi$ creation due to the dissociation of charmonia in the deconfinement phase of QCD \cite{Matsui_Satz}.
Suppression of $J/\psi$ particles was actually observed in several experiments at the Super Proton Synchrotron \cite{SPS} and the Relativistic Heavy Ion Collider \cite{RHIC}.
Here, not only $J/\psi$ but also $\chi_c$ and $\psi'$ states contribute to the total yield of $J/\psi$ \cite{E705}.
To investigate properties of these charmonia in the deconfinement phase, hadronic spectral functions are studied at zero and finite temperatures.

spectral functions on the lattice are calculated conventionally by using the maximum entropy method (MEM) \cite{MEM} 
in which spectral functions are extracted from Euclidean correlation functions using a Bayesian probability theory. 
With the MEM, the temperature dependence of charmonia spectral functions
has been investigated in quenched \cite{lqcd1,lqcd2,lqcd3,lqcd4} and two-flavor QCD \cite{lqcd5}.
From these studies, the S-wave charmonia ($\eta_c$, $J/\psi$) are suggested to survive up to temperatures higher than about 1.5$T_c$,
where $T_c$ is the critical temperature.
On the other hand, the P-wave charmonia ($\chi_{c0}$, $\chi_{c1}$) are suggested to dissolve just above $T_c$ \cite{lqcd4,lqcd5}.

The spectral functions from the MEM are continuous as expected in infinite-volume field theories.
In the finite volume,  on the other hand, the spectral functions consist of discrete spectra only, 
since the degrees of freedom of the theory is finite. 
Therefore, the meaning of the continuous spectral functions from the MEM is not quite clear.
This is problematic, in particular, in cases when the MEM leads to ambiguous results depending on, e.g., the choice of the default model.

In this paper, we propose a method to directly calculate spectral functions at discrete spectra on the lattice, applying a variational method.
Variational method is a powerful tool to extract information of low-lying discrete spectra \cite{var}.
In our previous paper, we studied the temperature dependence of charmonium wave functions for the ground and first excited states using a variational method \cite{ohno_umeda_kanaya}.
We extend the method to evaluate spectral functions, i.e., the location and the height for each low-lying discrete spectra.

With the variational method, information for low-lying spectra is extracted by diagonalizing correlation matrices between various smeared operators.
The spectral functions are defined in terms of correlation functions between pointlike operators.
We thus include pointlike operators among the smeared operators.
We calculate spectral functions from the element of the correlation matrices corresponding to the pointlike source and sink operators.
We show that
the accuracy and reliability of the resulting spectral functions can be systematically improved by increasing the number of smeared operators.

At nonzero temperatures, the spectral function for a meson operator $\mathcal{O}_{\Gamma}$ is given by
\begin{eqnarray}
\tilde\rho_\Gamma(\omega,\vec{p}) & = &
\sum_{k,k'} \frac{e^{-E_{k'}/T}}{Z(T)} \left(1-e^{-\omega / T}\right)
\left| \langle k' | \mathcal{O}_{\Gamma} | k \rangle \right|^2 
\nonumber\\
& \times &
(2\pi)^3 \delta(E_k-E_{k'}-\omega) \, \delta(\vec{p}_k-\vec{p}_{k'}-\vec{p}) ,
\label{eq:FTSPF}
\end{eqnarray}
where $E_k$ is the energy of the state $|k\rangle$ and $Z(T)$ is the finite temperature partition function.
As $T$ is increased, we expect that the height of each pole varies and, simultaneously, more poles at $E_k - E_{k'}$ appear in addition to the zero-temperature poles at $E_k - E_0$, where $E_0$ is the ground-state energy.
When a particle dissociates, we may expect that the corresponding spectral function shows a drastic change around the dissociation temperature:
The height of the pole corresponding to that particle will become lower and will be eventually buried by nearby poles.
Our final goal is to clarify the fate of various charmonia at high temperatures.

This paper is organized as follows.
We introduce our method to calculate locations and heights of the peaks of meson spectral functions with the variational method in the next section.
In Sec.~\ref{sec:free}, we first test the method for the case of free Wilson quarks. 
We construct meson correlation matrices using various Gaussian smearing functions and compare the results of spectral functions with analytic solutions.
We then apply the method to calculate the charmonium spectral functions in quenched QCD at zero and finite temperatures in Sec.~\ref{sec:results}. 
Our conclusions are given in Sec.~\ref{sec:conclusions}.
A preliminary report was presented in \cite{ohno_lat10}.

\section{\label{sec:method}Spectral functions with the variational method}

In this section, we introduce a method to calculate locations and heights of the peaks of meson spectral functions with a variational method \cite{var}.

The Euclidean meson correlation function is defined by
\begin{equation}\label{eq:meson_corr}
C_{\Gamma}(\vec{x},t) = \langle\mathcal{O}_{\Gamma}(\vec{x},t)\,\mathcal{O}^{\dag}_{\Gamma}(\vec{0},0)\rangle,
\end{equation}
where $\mathcal{O}_{\Gamma}(\vec{x},t) = \bar{q}(\vec{x},t)\Gamma q(\vec{x},t)$ is the pointlike meson operator for channel $\Gamma$ and
$\Gamma=\gamma_5,\gamma_i,\textrm{\boldmath $1$},\gamma_5\gamma_i$ $(i=1,2,3)$ 
correspond to pseudoscalar (Ps), vector (Ve), scalar (Sc) and axial-vector (Av) channels, respectively. 
For Ve and Av channels, we average the correlation functions over $i=1,2,3$.
Then, the meson spectral function $\tilde\rho_{\Gamma}(\omega,\vec{p})$ is related to its Fourier transform
\begin{equation}
C_{\Gamma}(\vec{p},t) = \int d^3x\, C_{\Gamma}(\vec{x},t)\; e^{i\vec{p}\cdot\vec{x}}
\end{equation}
by
\begin{equation}\label{eq:corr0}
C_{\Gamma}(\vec{p},t) = \int^{\infty}_{0} d\omega\, \tilde\rho_{\Gamma}(\omega,\vec{p})\,\frac{\cosh[\omega(t-\frac{1}{2T})]}{\sinh[\frac{\omega}{2T}]}
\end{equation}
where $T$ is the temperature. 
Setting the lattice spacing $a=1$, we have $1/T=N_t$ with $N_t$ the temporal lattice size.

On finite lattices, the spectral function consists of discrete spectra only.
Therefore, the integral over $\omega$ in (\ref{eq:corr0}) is actually a summation over discrete values of $\omega$.
For the zero momentum case, $\vec{p}=0$, we rewrite (\ref{eq:corr0}) on finite lattices as
\begin{equation}\label{eq:corr}
C_{\Gamma}(t) = \sum_{k} \rho_{\Gamma}(m_k)\,\frac{\cosh[m_k (t-N_t/2)]}{\sinh[m_k N_t/2]}
\end{equation}
with $k=1,2,\cdots$, where $\tilde\rho_\Gamma(\omega,\vec 0) = \sum_k \rho_\Gamma(m_k)\delta(\omega-m_k)$.

\begin{table}[tbh]
\caption{Smearing parameters $A_i$ for the test of the variational method. 
$A_1$ corresponds to the point operator.}
\label{smearing_param}
\begin{ruledtabular}
\begin{tabular}{cccccccc}
$A_1$  & $A_2$ & $A_3$ & $A_4$ & $A_5$ & $A_6$ & $A_7$ \\ \hline
$\infty$ & 0.25  & 0.20  & 0.15  & 0.10  & 0.05  & 0.02  \\
\end{tabular}
\end{ruledtabular}
\end{table}

\begin{figure}[tbh]
 \begin{center}
  \includegraphics[width=42mm, angle=-90]{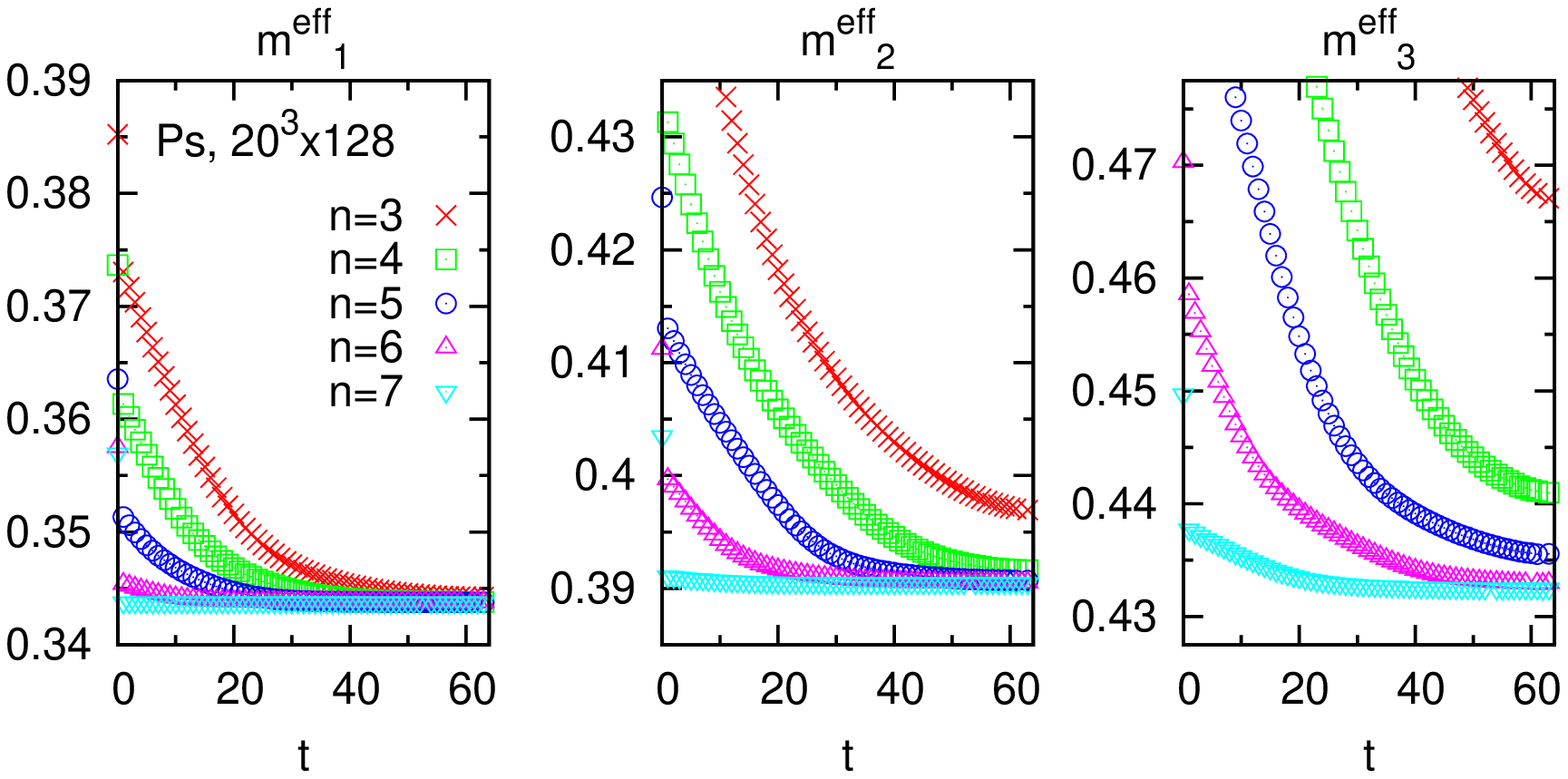}
  \includegraphics[width=42mm, angle=-90]{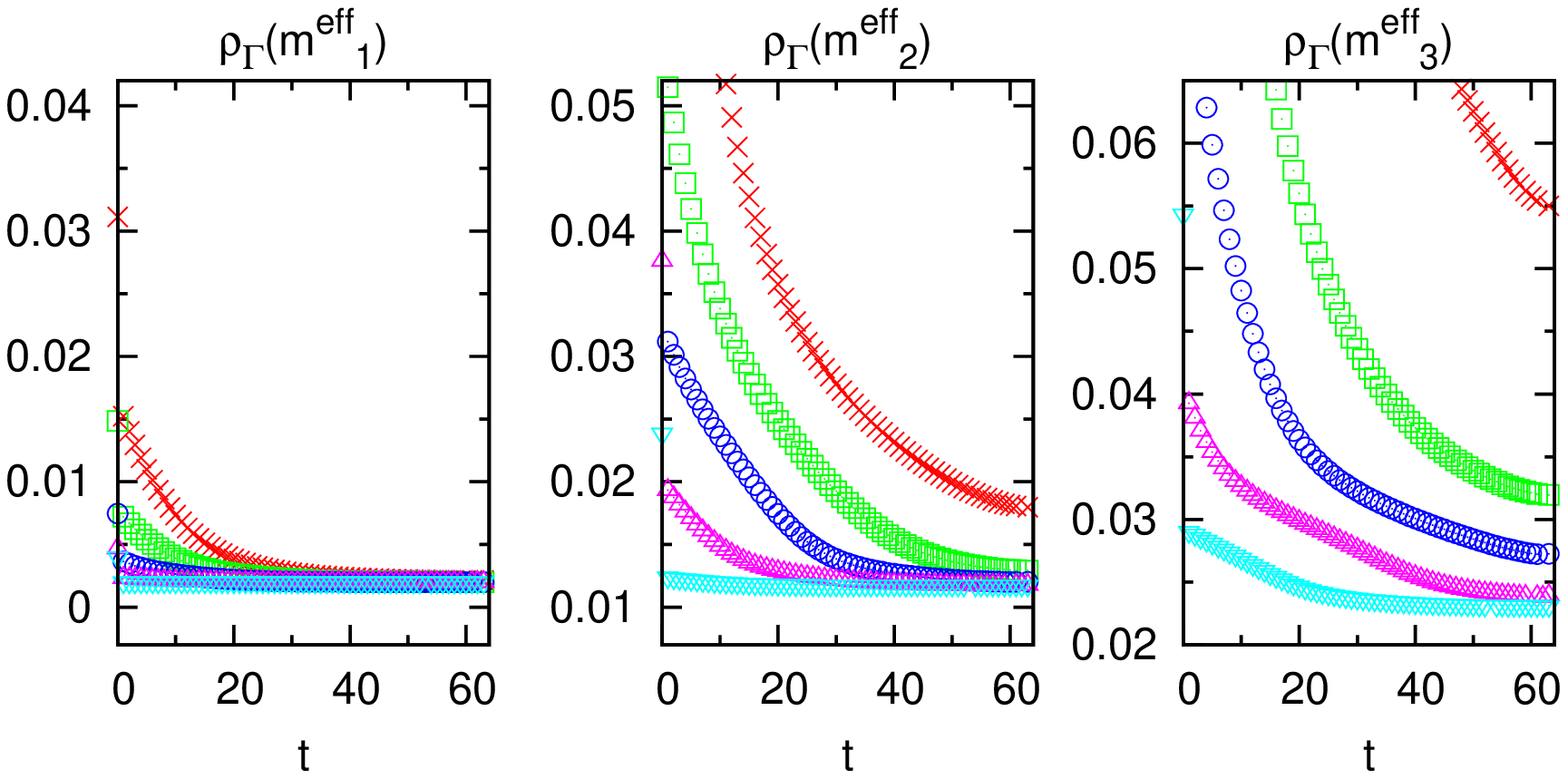}
  \vspace{-3mm}
  \caption{$t$- and $n$-dependence of $m^{\mathrm{eff}}_k$ (top) and $\rho_{\Gamma}(m^{\mathrm{eff}}_k)$ (bottom) obtained on the $20^3 \times 128$ lattice for the Ps channel.} 
  \label{t_dep_Ps:20128}
 \end{center}
\end{figure}
\begin{figure}[tbh]
 \begin{center}
  \includegraphics[width=42mm, angle=-90]{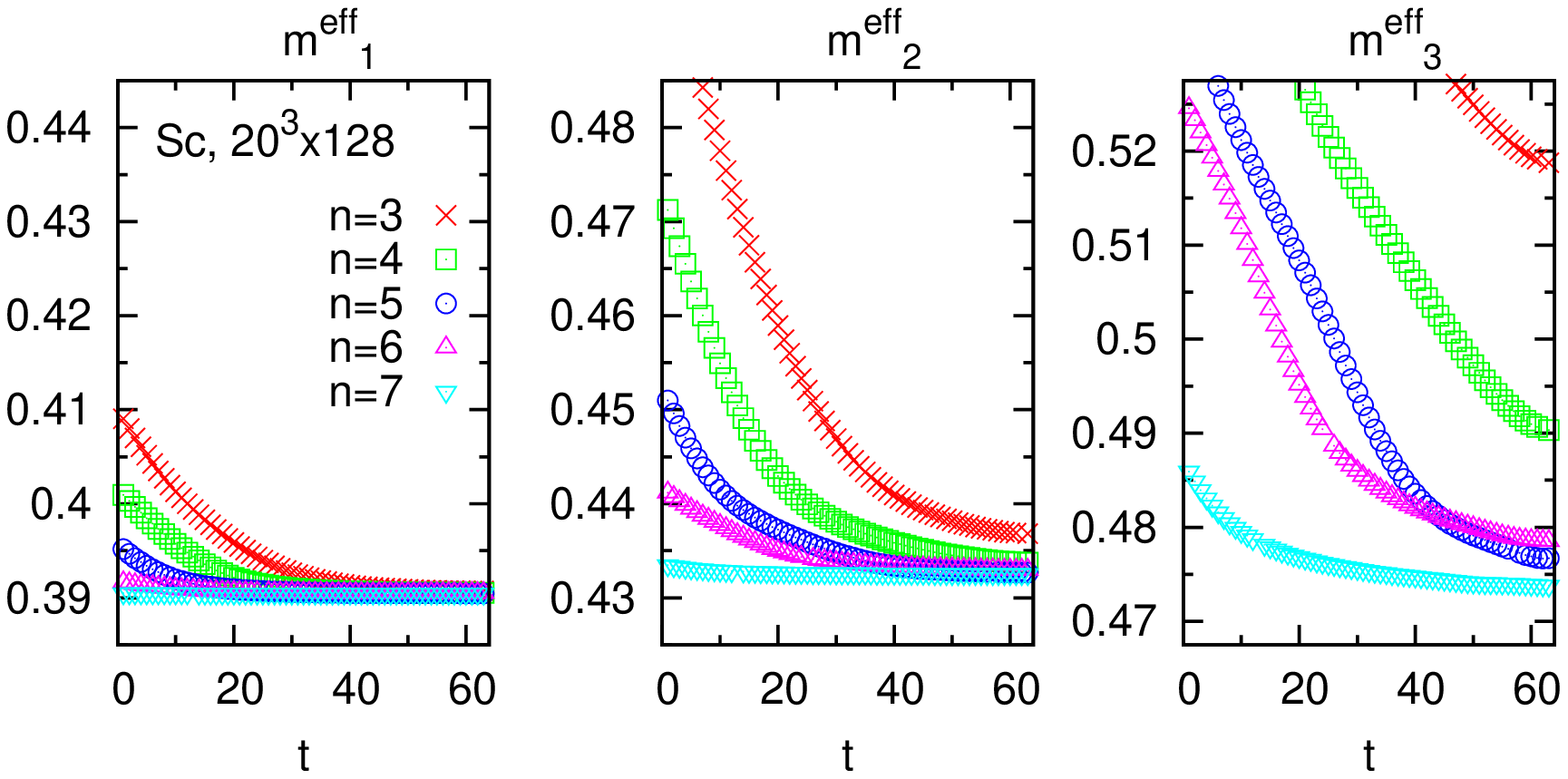}
  \includegraphics[width=42mm, angle=-90]{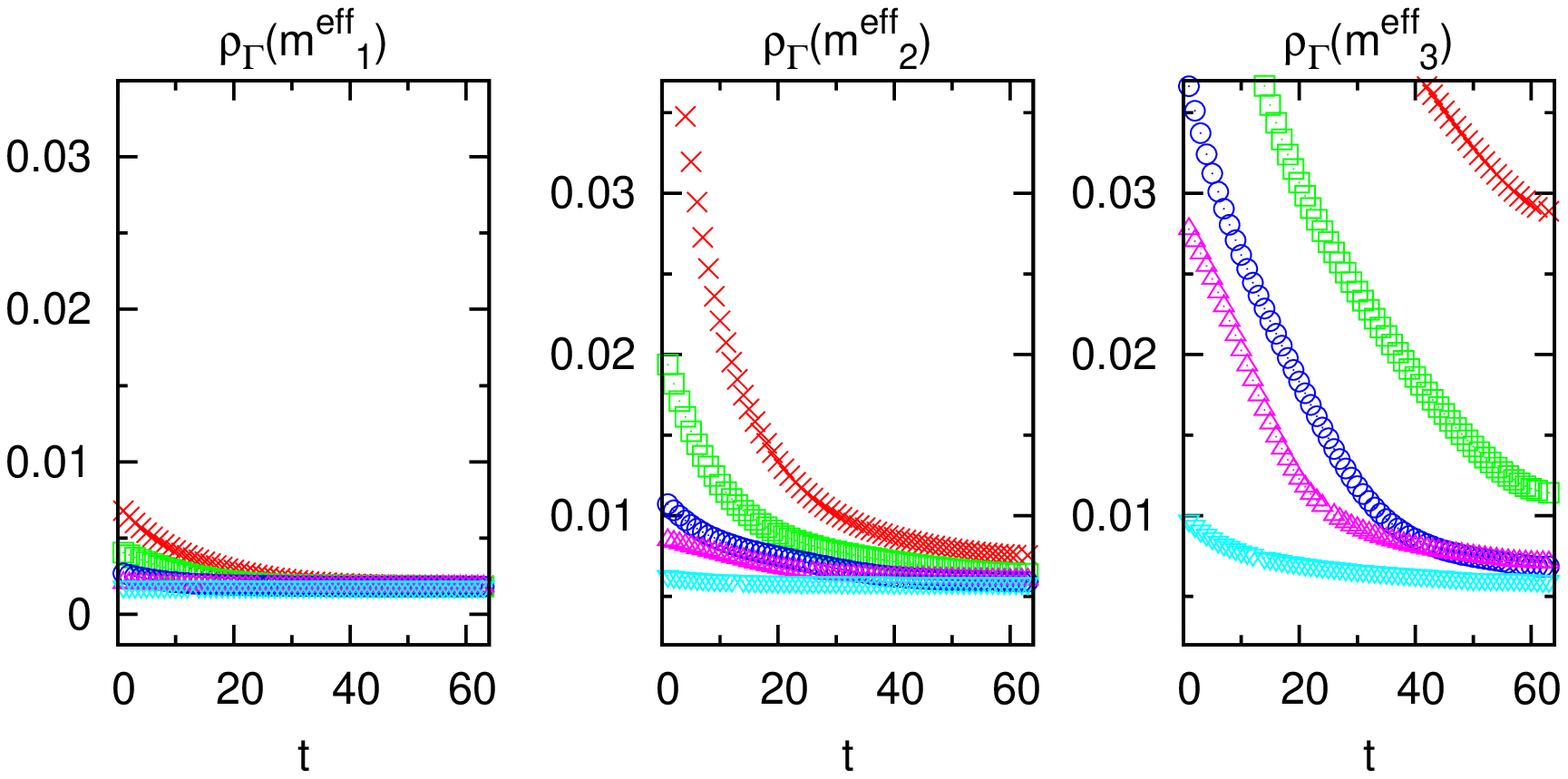}
  \caption{The same as Fig.~\ref{t_dep_Ps:20128} for the Sc channel.} 
  \label{t_dep_Sc:20128}
 \end{center}
\end{figure}

\begin{figure}[tbh]
 \begin{center}
  \includegraphics[width=57mm, angle=-90]{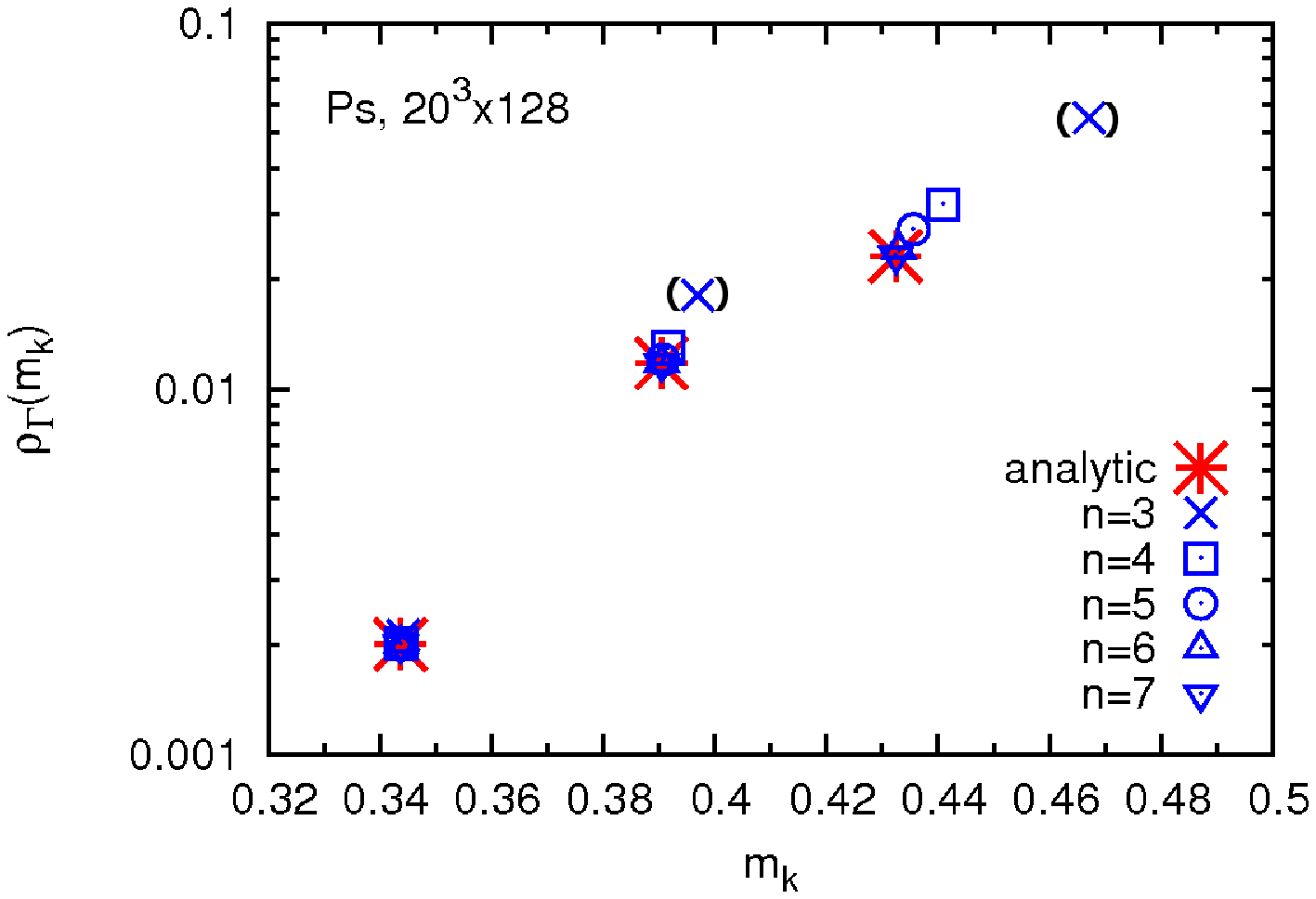}
  \includegraphics[width=57mm, angle=-90]{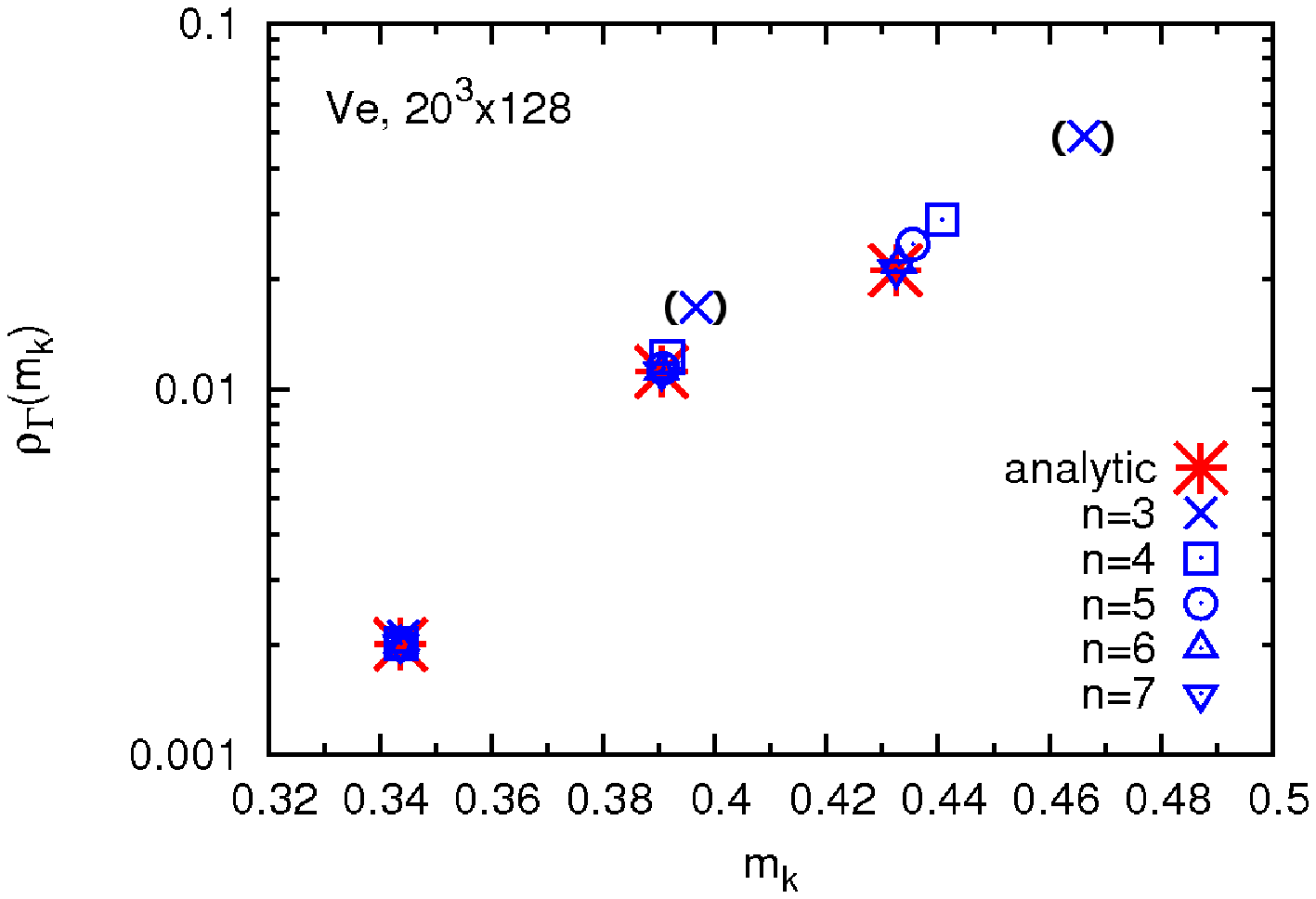}
  \vspace{-3mm}
  \caption{Meson spectral function at three lowest-lying states in the Ps and Ve channels obtained at $t=63$ on a $20^3\times 128$ lattice in the free quark case. 
  Cross, square, circle, triangle, and downward triangle symbols are the results of the variational method for $n=3,4,5,6,7$, respectively. 
  Symbols in the bracket mean that the corresponding signals are not asymptotic even at $t=63$.
  The analytic solutions are shown by the asterisks.} 
  \label{free_var_Ps}
 \end{center}
\end{figure}
\begin{figure}[tbh]
 \begin{center}
  \includegraphics[width=57mm, angle=-90]{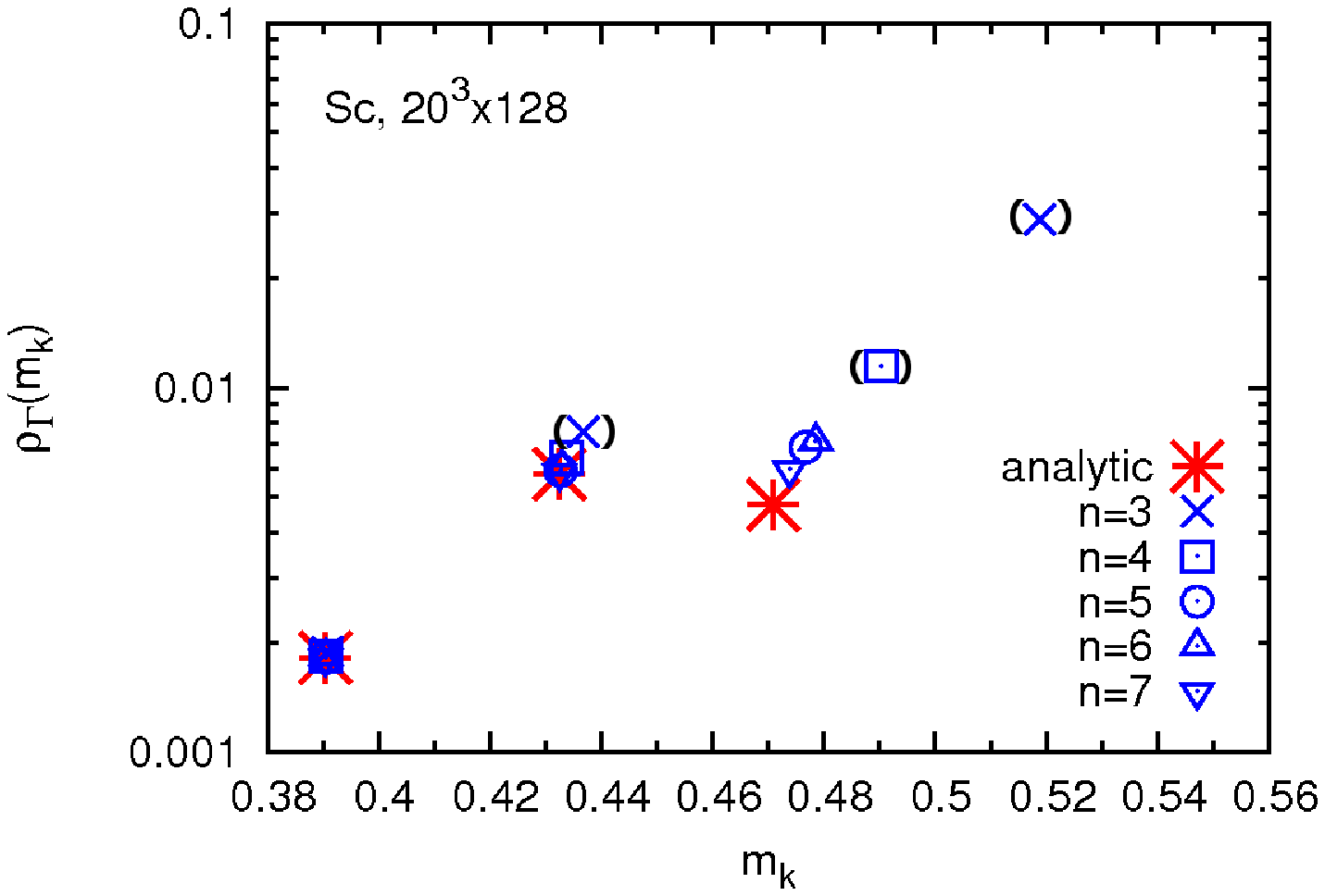}
  \includegraphics[width=57mm, angle=-90]{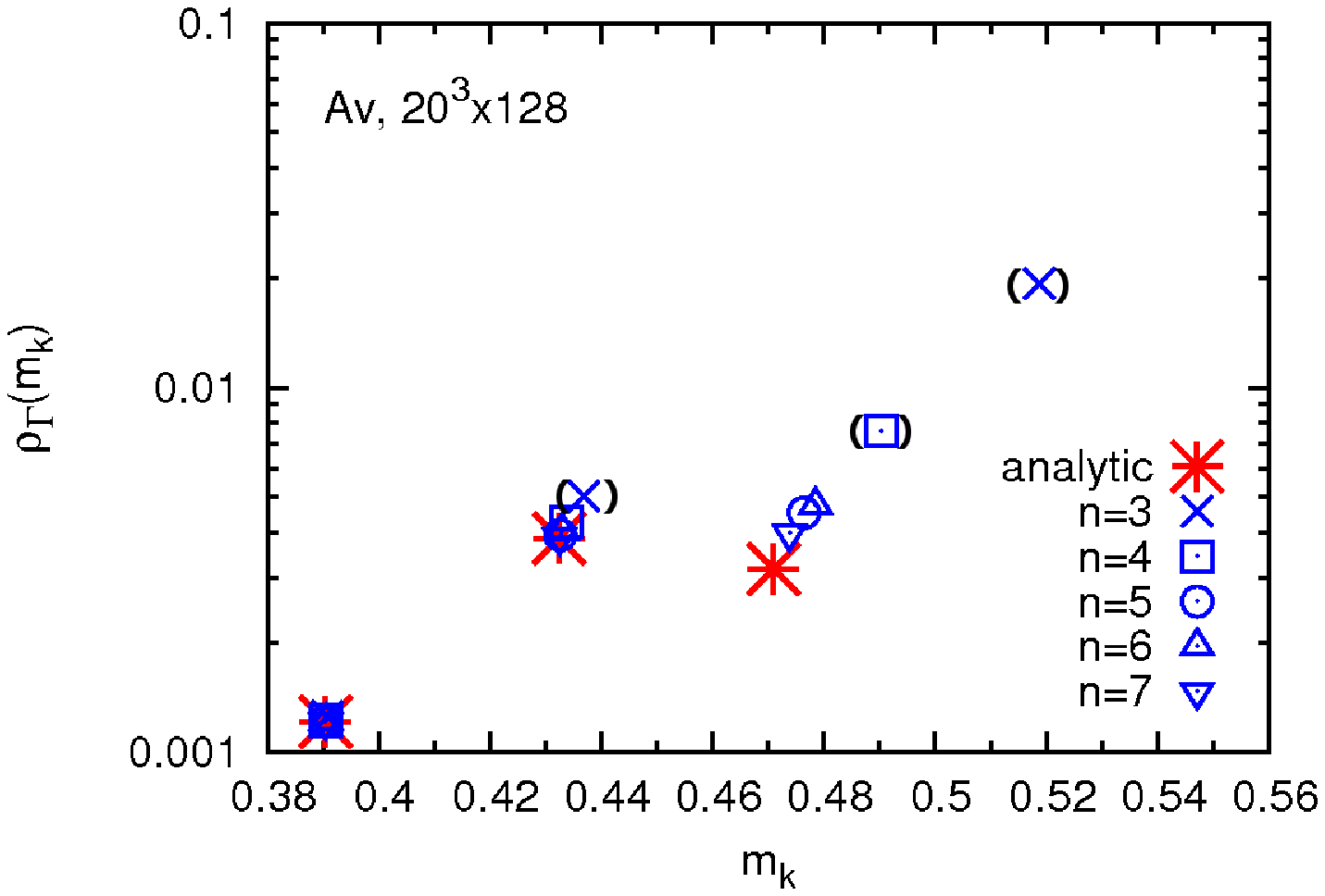}
  \vspace{-3mm}
  \caption{The same as FIG. \ref{free_var_Ps} for the Sc and Av channels.} 
  \label{free_var_Sc}
 \end{center}
\end{figure}

\begin{figure}[tbh]
 \begin{center}
  \includegraphics[width=42mm, angle=-90]{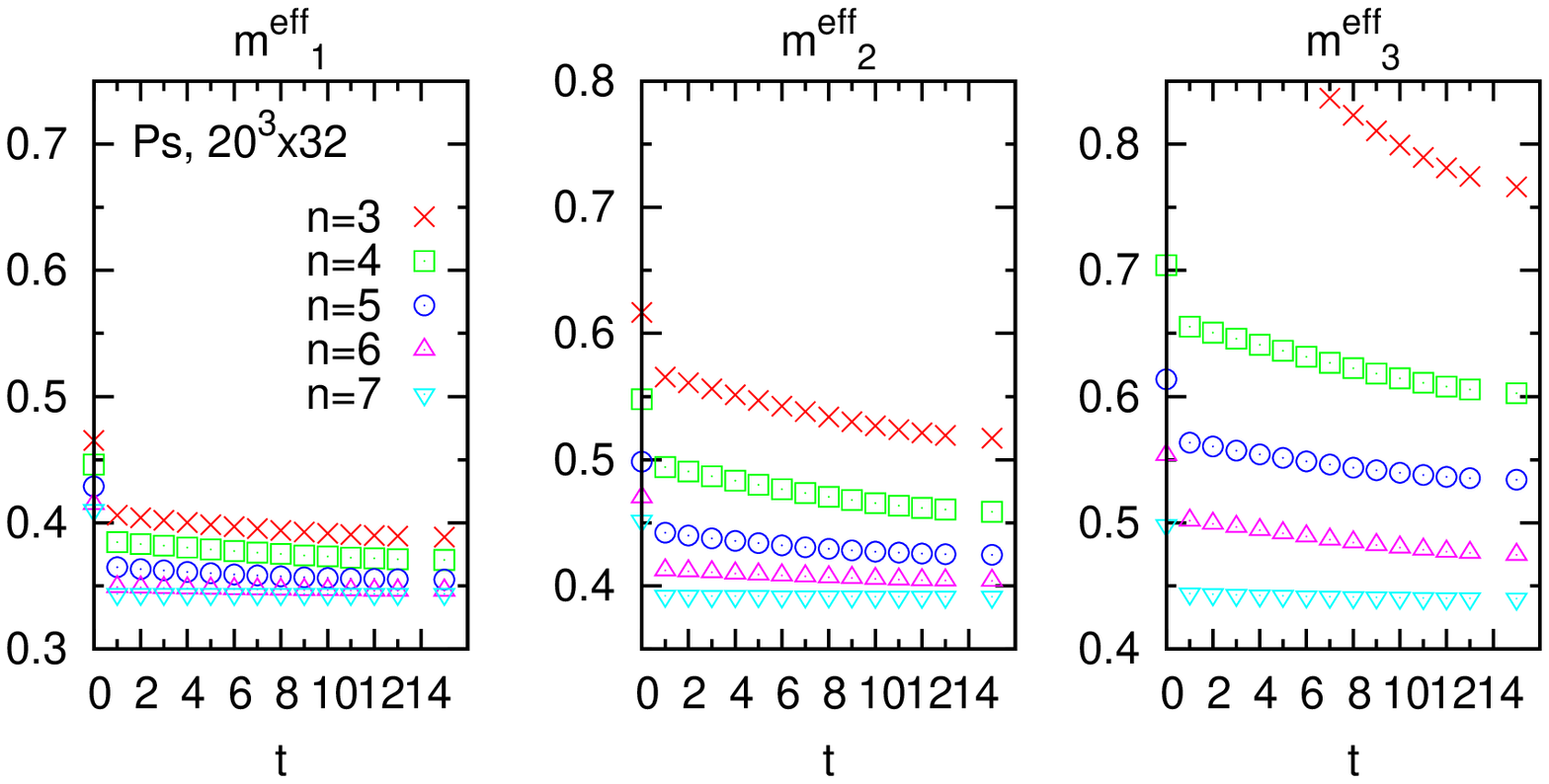}
  \includegraphics[width=42mm, angle=-90]{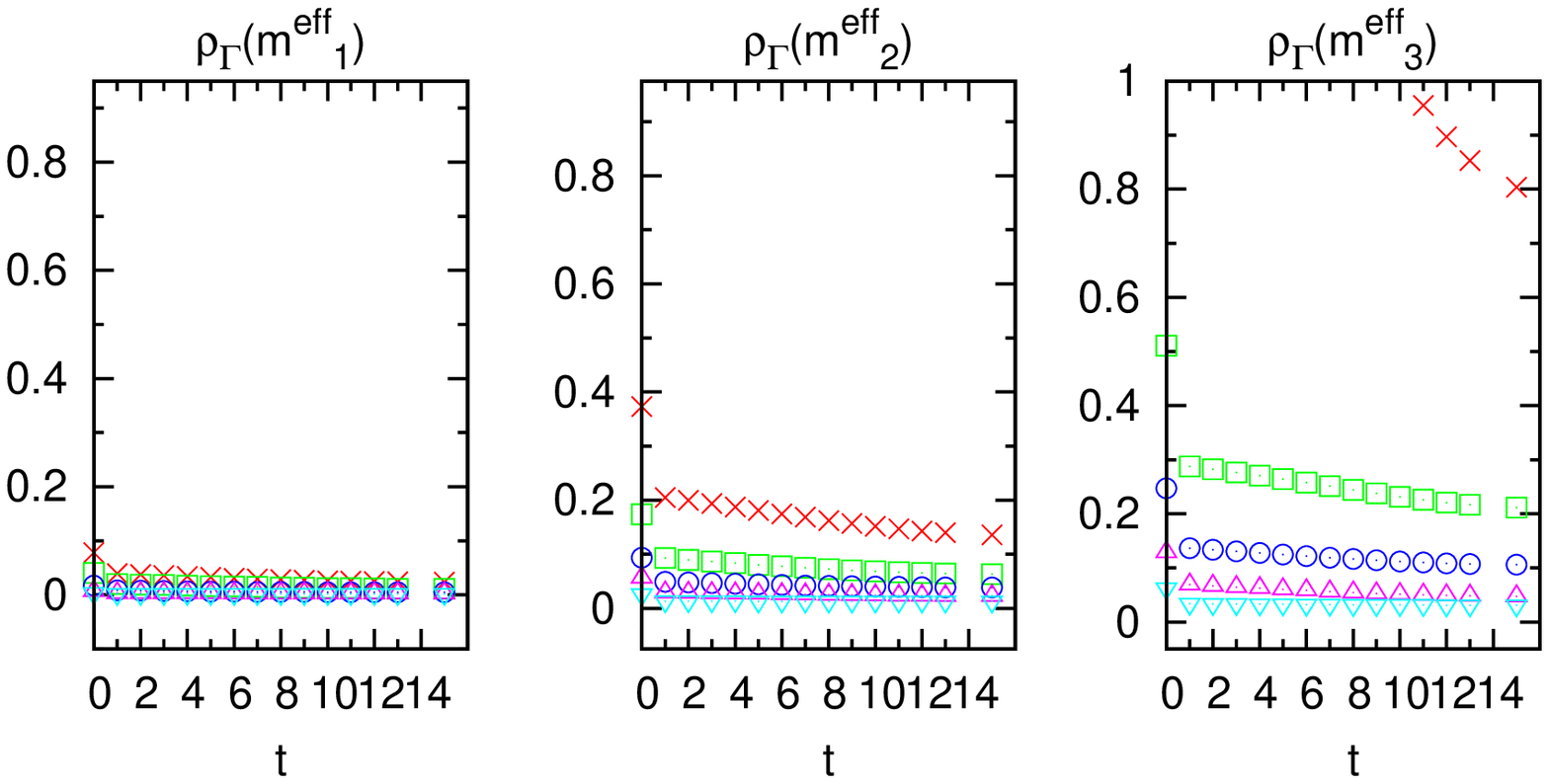}
  \caption{$t$- and $n$-dependence of $m^{\mathrm{eff}}_k$ (top) and $\rho_{\Gamma}(m^{\mathrm{eff}}_k)$ (bottom) obtained on the $20^3 \times 32$ lattice for the Ps channel.} 
  \label{t_dep_Ps:20032}
 \end{center}
\end{figure}
\begin{figure}[tbh]
 \begin{center}
  \includegraphics[width=42mm, angle=-90]{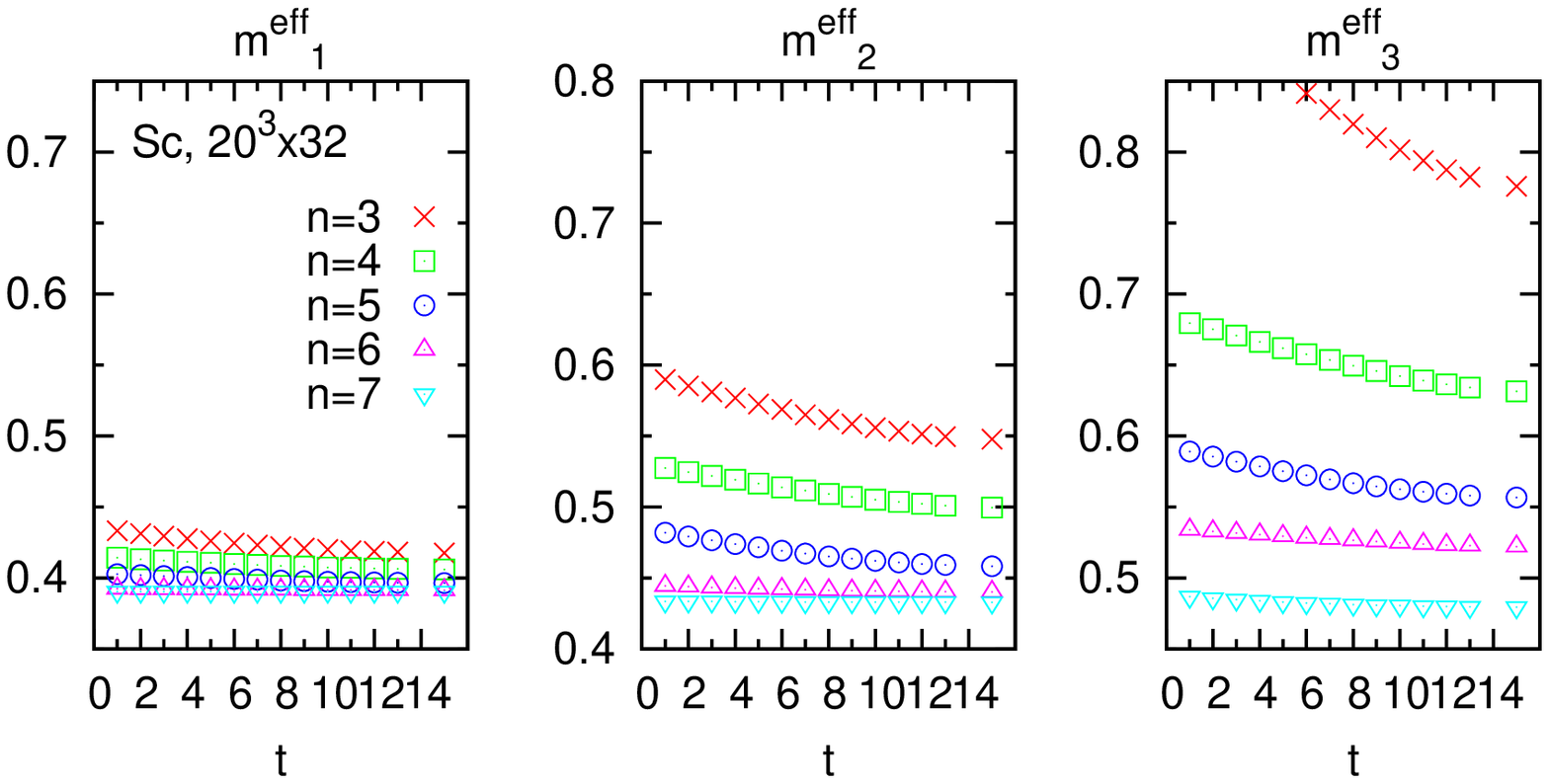}
  \includegraphics[width=42mm, angle=-90]{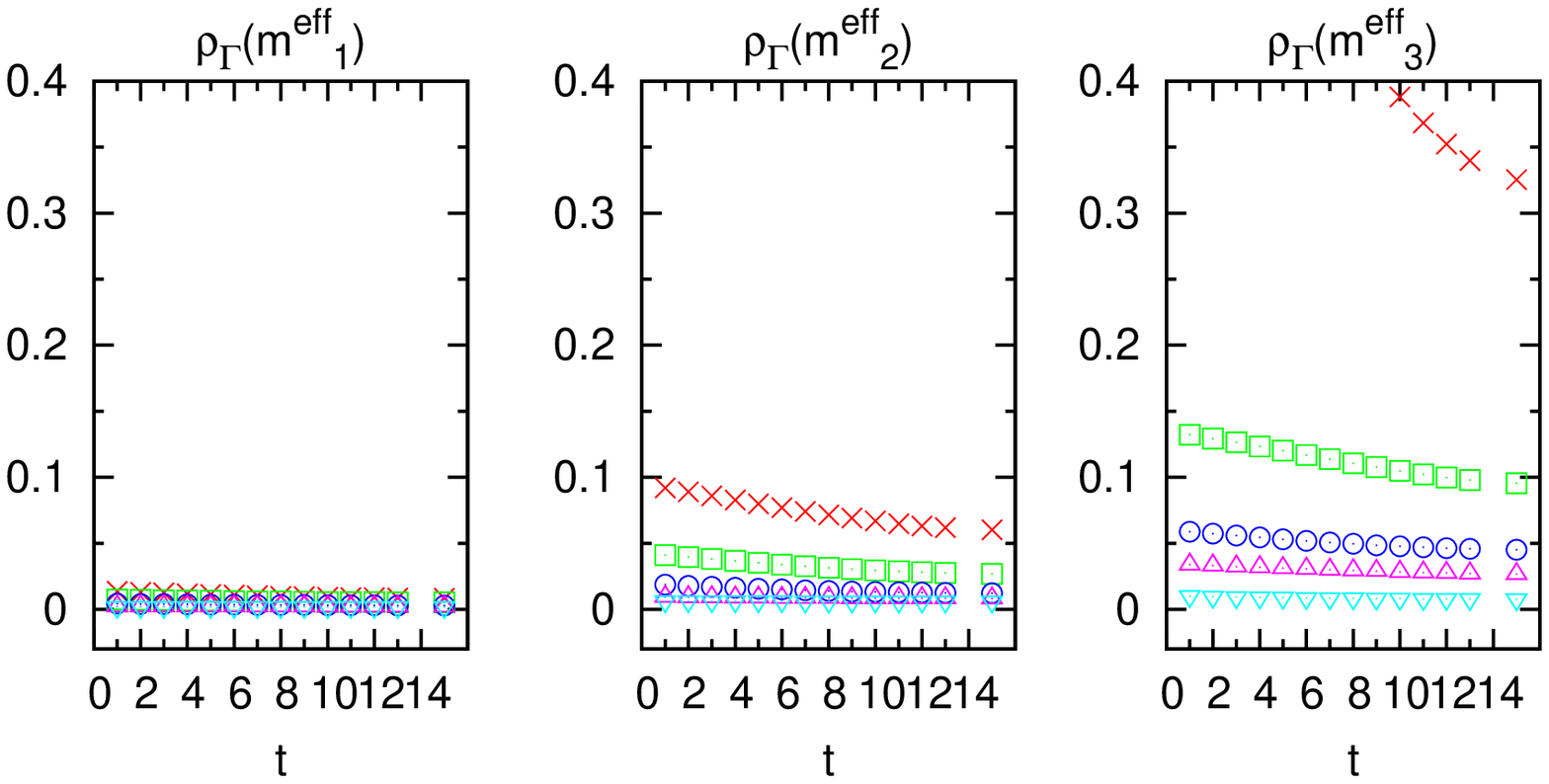}
  \caption{The same as Fig.~\ref{t_dep_Ps:20032} for the Sc channel.} 
  \label{t_dep_Sc:20032}
 \end{center}
\end{figure}

\begin{figure}[tbh]
 \begin{center}
  \includegraphics[width=57mm, angle=-90]{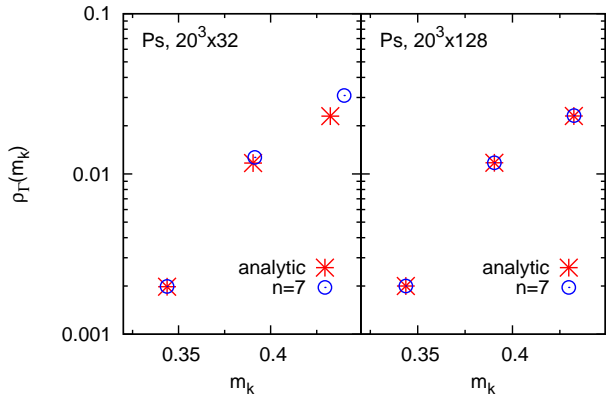}
  \vspace{-3mm}
  \caption{Meson spectral function at three lowest-lying states in the Ps channel obtained by the variational method with $n=7$
  at $t=15$ on the $20^3\times 32$ lattice (left), and at $t=63$ on the $20^3\times 128$ lattice (right)  in the free quark case. 
  The asterisk symbol is for the analytic solutions. }
  \label{free_Nt_Ps}
 \end{center}
\end{figure}
\begin{figure}[tbh]
 \begin{center}
  \includegraphics[width=57mm, angle=-90]{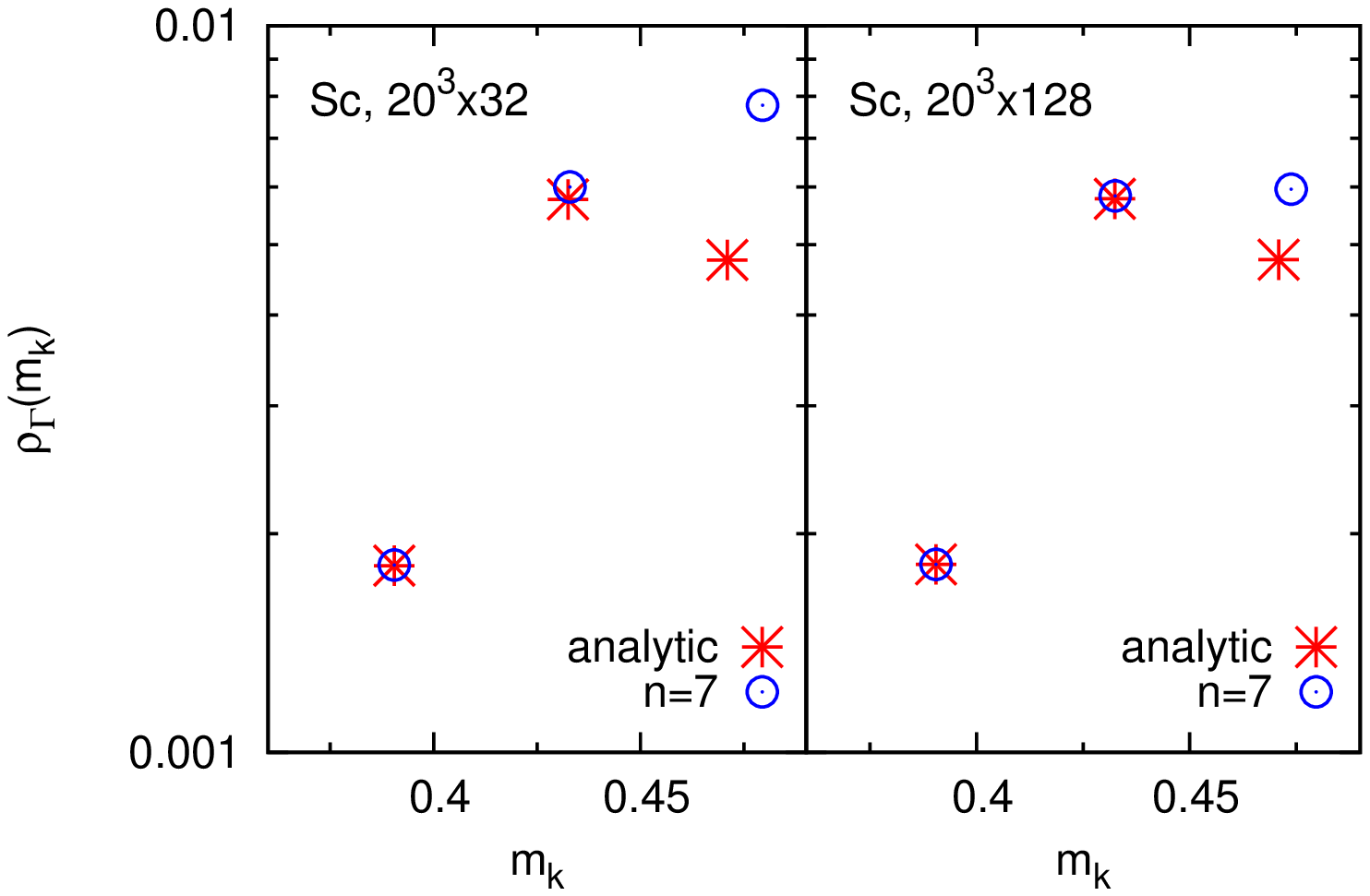}
  \vspace{-3mm}
  \caption{The same as Fig.~\ref{free_Nt_Ps} for the Sc channel.} 
  \label{free_Nt_Sc}
 \end{center}
\end{figure}

Let us introduce smeared meson operators 
\begin{equation}
\mathcal{O}_{\Gamma}(\vec{x},t)_i = \sum_{\vec{y},\vec{z}}\omega_i(\vec{y})\,\omega_i(\vec{z})\,\bar{q}(\vec{x}+\vec{y},t)\,\Gamma \,q(\vec{x}+\vec{z},t)
\end{equation}
with an appropriate gauge fixing \footnote{An alternative way to
adopt a gauge invariant smearing \cite{Allton:1993wc} for meson
operators. We leave studies in this direction for the future.},
where $\omega_i(\vec{x})$ $(i=1,2,\cdots,n)$ are smearing functions. 
We define an $n\times n$ meson correlation matrix 
${\bf C}_\Gamma(t) = \left[ C_\Gamma(t)_{i,j} \right]$
by
\begin{equation}
C_{\Gamma}(t)_{ij}=\sum_{\vec{x}}\langle\mathcal{O}_{\Gamma}(\vec{x},t)_i\,\mathcal{O}^{\dag}_{\Gamma}(\vec{0},0)_j\rangle.
\end{equation}
In this study, we set $\omega_{1}(\vec{x}) = \delta(\vec{x})$, so that $C_{\Gamma}(t)_{11}$ is only the $C_{\Gamma}(t)$ defined by (\ref{eq:corr}).

By solving a generalized eigenvalue problem 
\begin{equation}\label{eq:gep}
{\bf C}_{\Gamma}(t)\,{\bf v}^{(k)} = \lambda_k(t;t_0)\, {\bf C}_{\Gamma}(t_0)\,{\bf v}^{(k)}
\end{equation}
for $k=1,\cdots,n$, we define effective masses $m^{\mathrm{eff}}_k(t;t_0)$ by
\begin{equation}\label{eq:lambda}
\lambda_k(t;t_0) = \frac{\cosh[m^{\mathrm{eff}}_k(t;t_0)(t-N_t/2)]}{\cosh[m^{\mathrm{eff}}_k(t;t_0)(t_0-N_t/2)]}.
\end{equation}
Denoting $\Lambda = {\rm diag}\{\lambda_1,\cdots,\lambda_n\}$ and
${\bf V} = [{\bf v}^{(1)} \cdots {\bf v}^{(n)}]$,
we rewrite (\ref{eq:gep}) as ${\bf C}_{\Gamma}(t) = {\bf C}_{\Gamma}(t_0) {\bf V} \Lambda {\bf V}^{-1}$.
Then, the (1,1) element of this relation reads
\begin{eqnarray}\label{eq:corr2}
&& C_{\Gamma}(t)_{11} \nonumber \\
&=& \sum_{k}\left({\bf C}_{\Gamma}(t_0){\bf V}\right)_{1k}({\bf V}^{-1})_{k1}\frac{\sinh[m^{\mathrm{eff}}_k(t;t_0)N_t/2]}{\cosh[m^{\mathrm{eff}}_k(t;t_0)(t_0-N_t/2)]} \nonumber \\ 
&& \times \frac{\cosh[m^{\mathrm{eff}}_k(t;t_0)(t-N_t/2)]}{\sinh[m^{\mathrm{eff}}_k(t;t_0)N_t/2]}.
\end{eqnarray}
Comparing (\ref{eq:corr}) and (\ref{eq:corr2}), we define an effective spectral function
\begin{eqnarray}\label{eq:final}
\rho_{\Gamma}(m^{\mathrm{eff}}_k(t;t_0)) &=& \left({\bf C}_{\Gamma}(t_0){\bf V}\right)_{1k}({\bf V}^{-1})_{k1} \nonumber \\
&\times& \frac{\sinh[m^{\mathrm{eff}}_k(t;t_0)\,N_t/2]}{\cosh[m^{\mathrm{eff}}_k(t;t_0)\,(t_0-N_t/2)]}.
\end{eqnarray}
When we let $t$ and $t_0$ sufficiently large, $m^{\mathrm{eff}}_k$ approaches to the mass of the $k$-th state $m_k$ and
$\rho_\Gamma(m^{\mathrm{eff}}_k)$ approaches to the spectral function $\rho_\Gamma(m_k)$ defined by (\ref{eq:corr}).

Note that while only the (1,1) element is related to $\rho_{\Gamma}(m^{\mathrm{eff}}_k)$, all $n$ trial states contribute in (\ref{eq:final}).
Keeping $n$ finite introduces a systematic error in the location and height of low-lying spectra at finite $t$.
By increasing $n$, we can systematically improve the results. 
On the other hand, we note that setting $n$ too large can lead to large statistical fluctuations and numerical instabilities.

In the calculations shown below, we further apply the midpoint subtraction method \cite{const_mode} to the meson correlator matrices in order to subtract the constant mode contributions:
\begin{eqnarray}\label{eq:midpoint}
{\bf C}_{\Gamma}(t) \;\;\rightarrow\;\; {\bf C}_{\Gamma}(t) - {\bf C}_{\Gamma}(N_t/2).
\end{eqnarray}
Accordingly, $\cosh$ in 
(\ref{eq:corr}) and (\ref{eq:lambda})--(\ref{eq:final})  
should be modified as
$
\cosh\left[m (t-N_t/2)\right] \;\rightarrow\; \cosh\left[m (t-N_t/2)\right] - 1 
$
while other factors including the $\sinh$ terms remain unchanged.
In principle, we can treat the constant mode as an eigenstate in the variational method too.
We have confirmed that the low-lying physical modes from variational calculations with and without the midpoint subtraction procedure are consistent with each other. 
We find, however, that the midpoint subtraction improves the arithmetic precision of signals at large $t$ and thus the resulting effective spectral functions are more stable with the midpoint subtraction.

\section{\label{sec:free}Test with free quarks}

In order to test the method, we first study the case of free Wilson quarks.
We compare spectral functions from the variational method with the analytic solutions on anisotropic lattices with the anisotropy $\xi=a_s/a_t=4$, where $a_s$ and $a_t$ are the spatial and temporal lattice spacings, respectively.
At the Wilson parameter $r=1$, we adjust the quark mass $\hat{m} \approx 0.7501$ to approximately reproduce the grand-state meson masses in quenched QCD studied in the next section.

The analytic solutions for meson spectral functions with free Wilson quarks are given in Appendix \ref{sec:analytic_SPF}  for the Ps, Ve, Sc and Av channels. 
The spectral functions with the variational method are calculated by setting link variables to unity in the QCD code to be used in the next section.
In this paper, we adopt Gaussian smearing functions defined by
\begin{equation}
\omega_i(\vec{x}) = e^{-A_i |\vec{x}|^2},\quad i=1,2,\cdots,n,
\label{eq:Gsmearing}
\end{equation}
with the smearing parameters $A_i$ listed in Table~\ref{smearing_param}.
Here $A_1=\infty$ is for the point operator.

We first test on a $20^3 \times 128$ lattice with $\xi=4$. 
In Figs.~\ref{t_dep_Ps:20128} and \ref{t_dep_Sc:20128}, we show the effective mass $m^{\mathrm{eff}}_k(t,t_0)$ (top) and the effective spectral function $\rho_{\Gamma}(m^{\mathrm{eff}}_k(t,t_0))$ (bottom) for the lowest three states in the Ps and Sc channels as functions of $t$.
We extract the signals at the largest $t=63$ ($t=64$ is excluded by the midpoint subtraction procedure).
In order to suppress the contamination of higher states, we choose large $t_0$ $(< t)$ under the condition that the signals of $m^{\mathrm{eff}}_k$ and $\rho(m^{\mathrm{eff}}_k)$ for the lowest three states are stable for all values of $t$ up to $n=7$ (see Appendix~\ref{free_SPF}).
The values of $t_0$ are given in Appendix~\ref{free_SPF}.
The results for the Ve (Av) channel are similar to those of the Ps (Sc) channel. 

From these figures, we find that the ground-state signals $m_1$ and $\rho(m_1)$ can be safely extracted even with a small $n$.
A larger $n$ is required for excited states to obtain asymptotic signals at $t=63$,
in particular, for the second excited state ($k=3$) of P-waves.
On the other hand, we find that setting $n$ too large cause numerical instabilities due to the limitation of the arithmetic precision -- in the present test, $n>7$ induces instabilities at large $t$ for several states.
We thus restrict ourselves to $n \le 7$.

Figures \ref{free_var_Ps} and \ref{free_var_Sc} show the results of $m_k$ and $\rho_{\Gamma}(m_k)$ for the lowest three states obtained at $t=63$ with $n=3$, 4, $\cdots$, 7.
The results in the bracket are those apparently not asymptotic at $t=63$ from the $t$-dependence of the effective mass or the effective spectral function.
The analytic solutions are given by the asterisks.
We find that, except for the case of the second excited states in P-wave, the analytic solutions are well reproduced by choosing a sufficiently large $n$.
For the second excited states in P-wave, we observe slight deviations from the analytic results even with our largest $n$  (see Fig.~\ref{free_var_Sc}).
The fact that the deviations become smaller with increasing $n$ suggests that these results are not fully asymptotic yet.
A possible cause will be our choice of the trial states based on the Gaussian smearing functions which may not be overlapping well with P-wave excited states.
Another reason may be the large contamination of the constant mode in P-wave correlation functions (see Appendix~\ref{sec:analytic_SPF}).
Although the constant mode is removed by the midpoint subtraction procedure, the resulting signal suffers from lower precision.
We leave these issues for future investigations.

On finite temperature lattices with a small temporal extent $N_t$, the range of $t$ available for the variational analyses is limited. 
To study its influences, we repeat the test on an anisotropic $20^3 \times 32$ lattice with $\xi=4$, adopting the same simulation parameters.
In the free quark case, because no interactions with the thermal background medium exist, we have no additional poles at $T>0$ in (\ref{eq:FTSPF}).

Effective masses and effective spectral functions obtained on the finite temperature lattice are shown in Figs.~\ref{t_dep_Ps:20032} and \ref{t_dep_Sc:20032} for the Ps and Sc channels.
We adopt $t_0=14$ and $t=15$. 
In Figs.~\ref{free_Nt_Ps} and \ref{free_Nt_Sc}, we compare spectral functions on $20^3 \times 32$ and $20^3 \times 128$ lattices for the lowest three states in these channels obtained with $n=7$.
The asterisks represent the analytic solutions. 
Results for the Ve and Av channels are similar to those of the Ps and Sc channels, respectively. 
We find that the limitation of the range of $t$ requires a larger $n$ to extract asymptotic signals. 
On the $N_t=32$ lattice, the second excited states show deviations from the analytic results even with $n=7$.
To overcome the problem, a set of more optimally smeared operators will be needed.
By examining both $t$ and $n$ dependences of the results, however, we can get an idea to which extent the results are asymptotic.

\begin{table}[tb]
\caption{The simulation parameters for the plaquette gauge action and the $O(a)$-improved Wilson quark action on our anisotropic lattice.}
\label{sim_param}
\begin{ruledtabular}
\begin{tabular}{ccccccccc}
$\beta$ & $\xi$ & $\gamma_G$ & $\gamma_F$ & $r$ & $c_E$ & $c_B$ & $\kappa$ \\ \hline
6.10    & 4     & 3.2108     & 4.94       & 1   & 3.164 & 1.911 & 0.10109  \\
\end{tabular}
\end{ruledtabular}
\end{table}

\begin{figure}[tbh]
 \begin{center}
  \includegraphics[width=60mm, angle=-90]{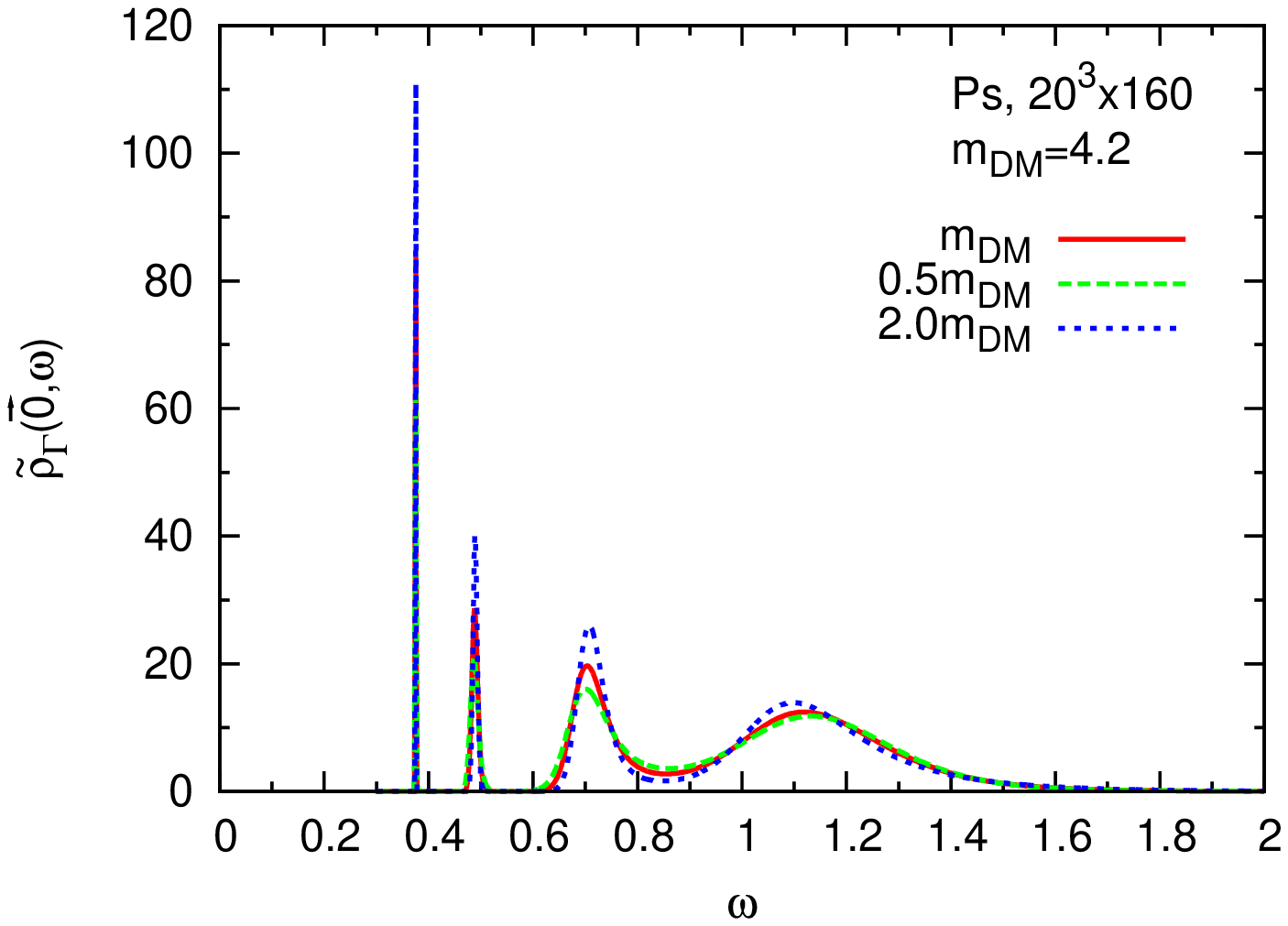}
  \includegraphics[width=60mm, angle=-90]{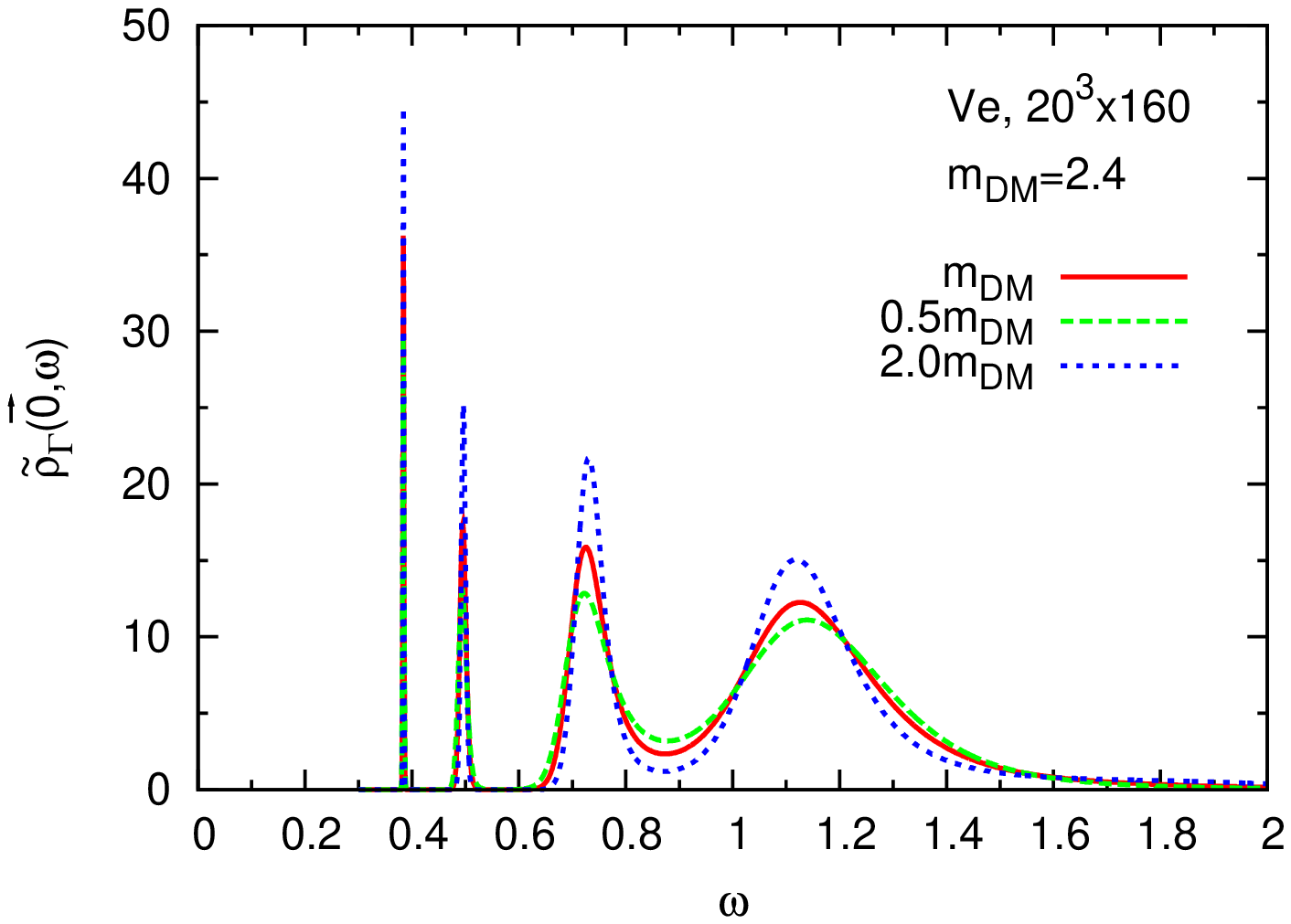}
  \caption{Charmonium spectral function calculated by MEM for the Ps and Ve channels at zero temperature. We use (\ref{eq:DM})
	as the default model function.
        The results for $\alpha=0.5,1.0,2.0$ are shown by solid, dashed, and dotted lines, respectively.} 
  \label{mem_Ps}
 \end{center}
\end{figure}
\begin{figure}[tbh]
 \begin{center}
  \includegraphics[width=60mm, angle=-90]{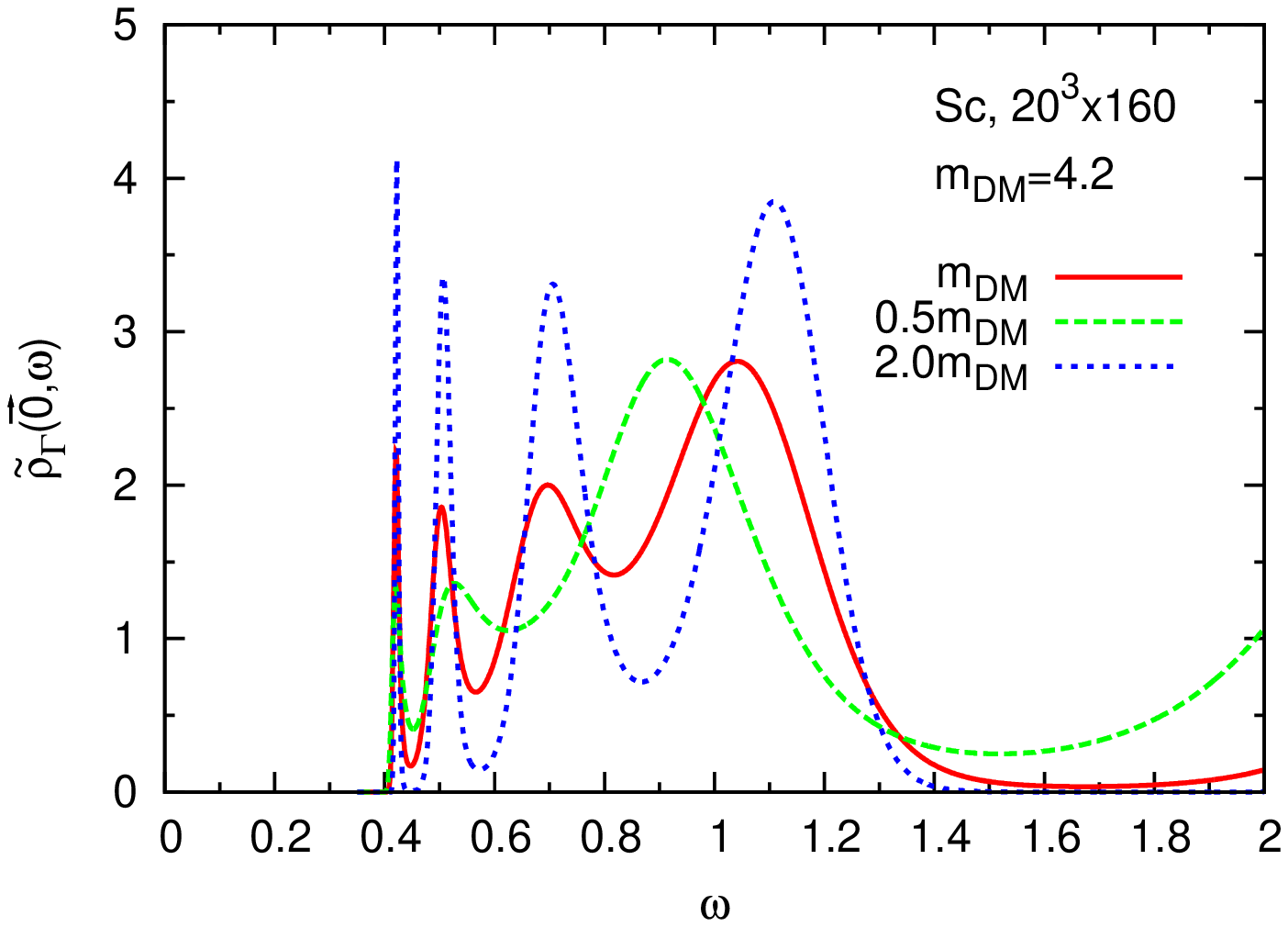}
  \includegraphics[width=60mm, angle=-90]{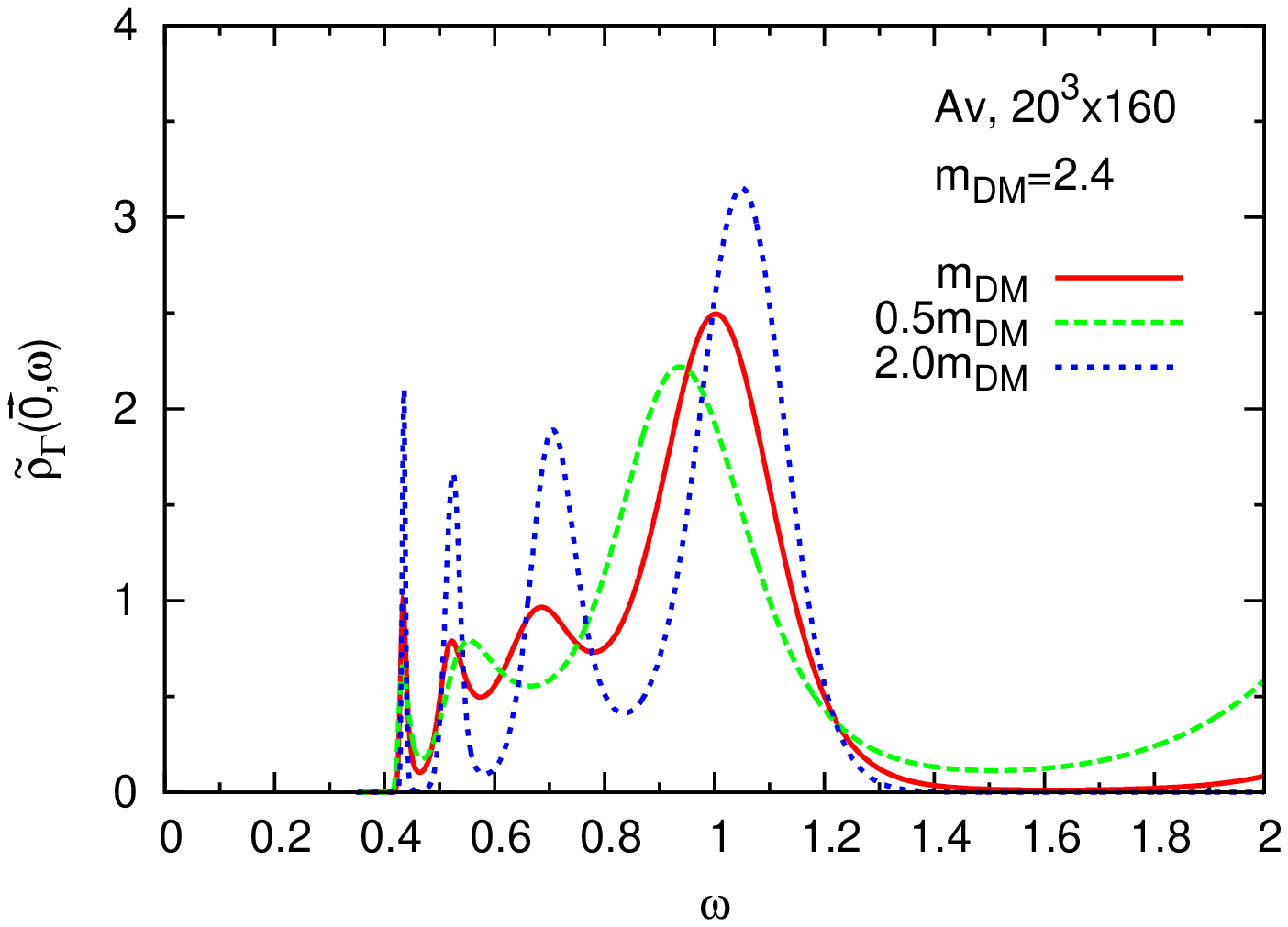}
  \caption{The same as Fig.~\ref{mem_Ps} for the Sc and Av channels.} 
  \label{mem_Sc}
 \end{center}
\end{figure} 

\begin{figure}[tbh]
 \begin{center}
  \includegraphics[width=44mm, angle=-90]{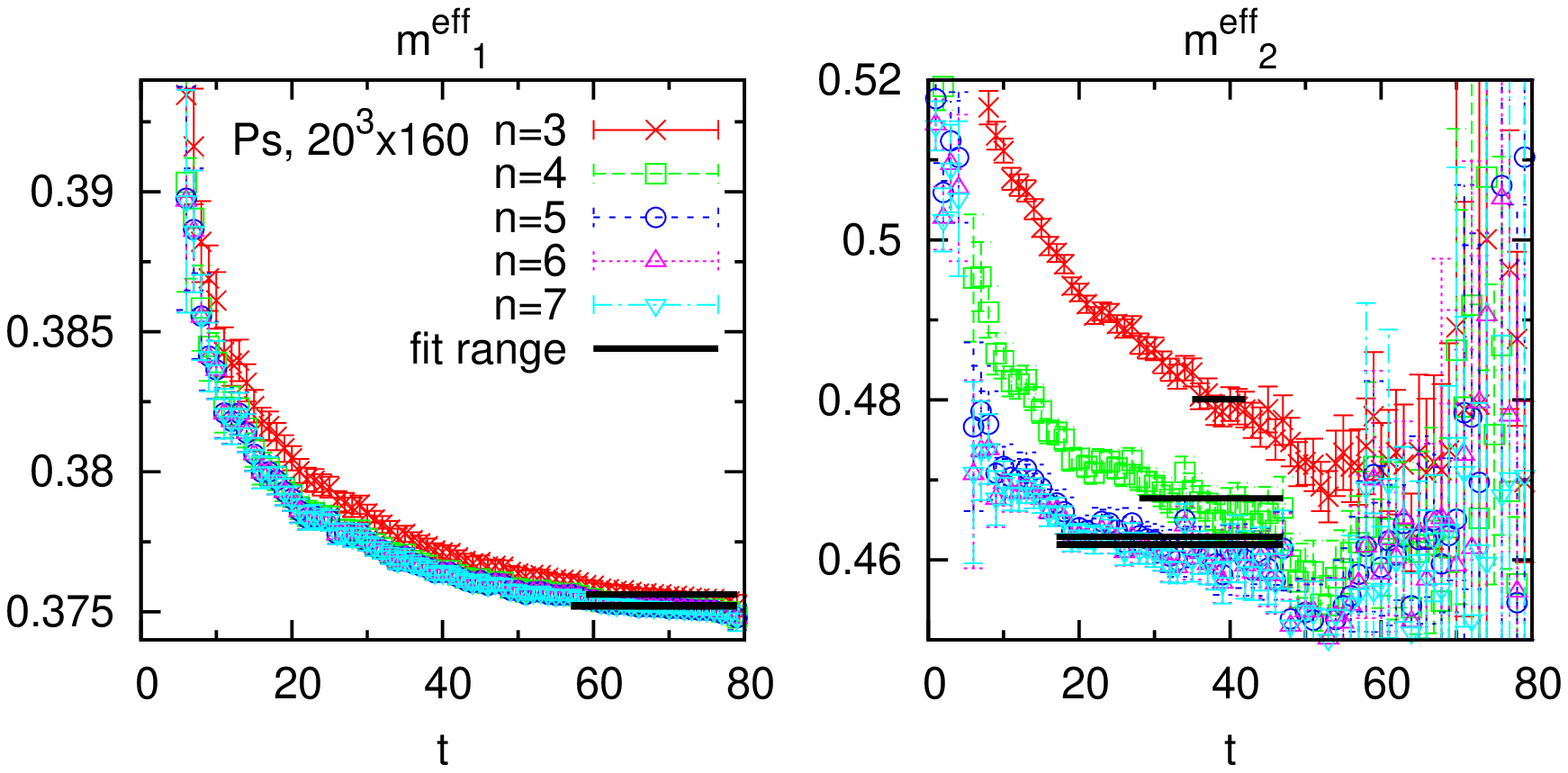}
  \includegraphics[width=44mm, angle=-90]{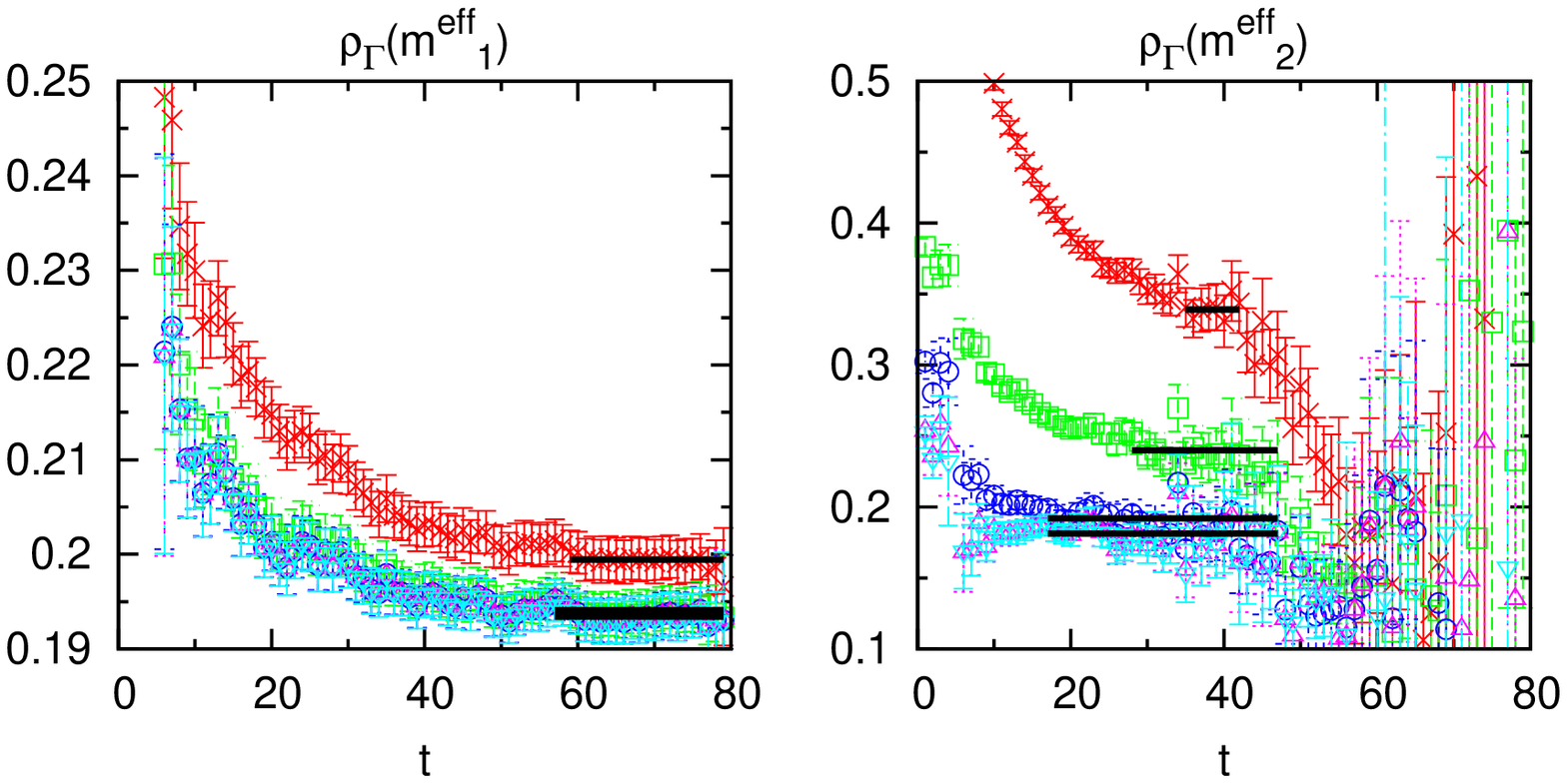}
  \caption{$m^{\mathrm{eff}}_1(t,t_0)$ (left-top), $m^{\mathrm{eff}}_2(t,t_0)$ (right-top), $\rho_{\Gamma}(m^{\mathrm{eff}}_1(t,t_0))$ (left-bottom) and $\rho_{\Gamma}(m^{\mathrm{eff}}_2(t,t_0))$ (right-bottom) of chamonium for Ps channel
  at zero temperature with the variational method. The reference point is chosen at $t_0=5$. Fit ranges to determine the locations and heights of the spectral peaks for charmonia are also shown by horizontal solid lines.} 
  \label{fit_Ps}
 \end{center}
\end{figure}
\begin{figure}[tbh]
 \begin{center}
  \includegraphics[width=44mm, angle=-90]{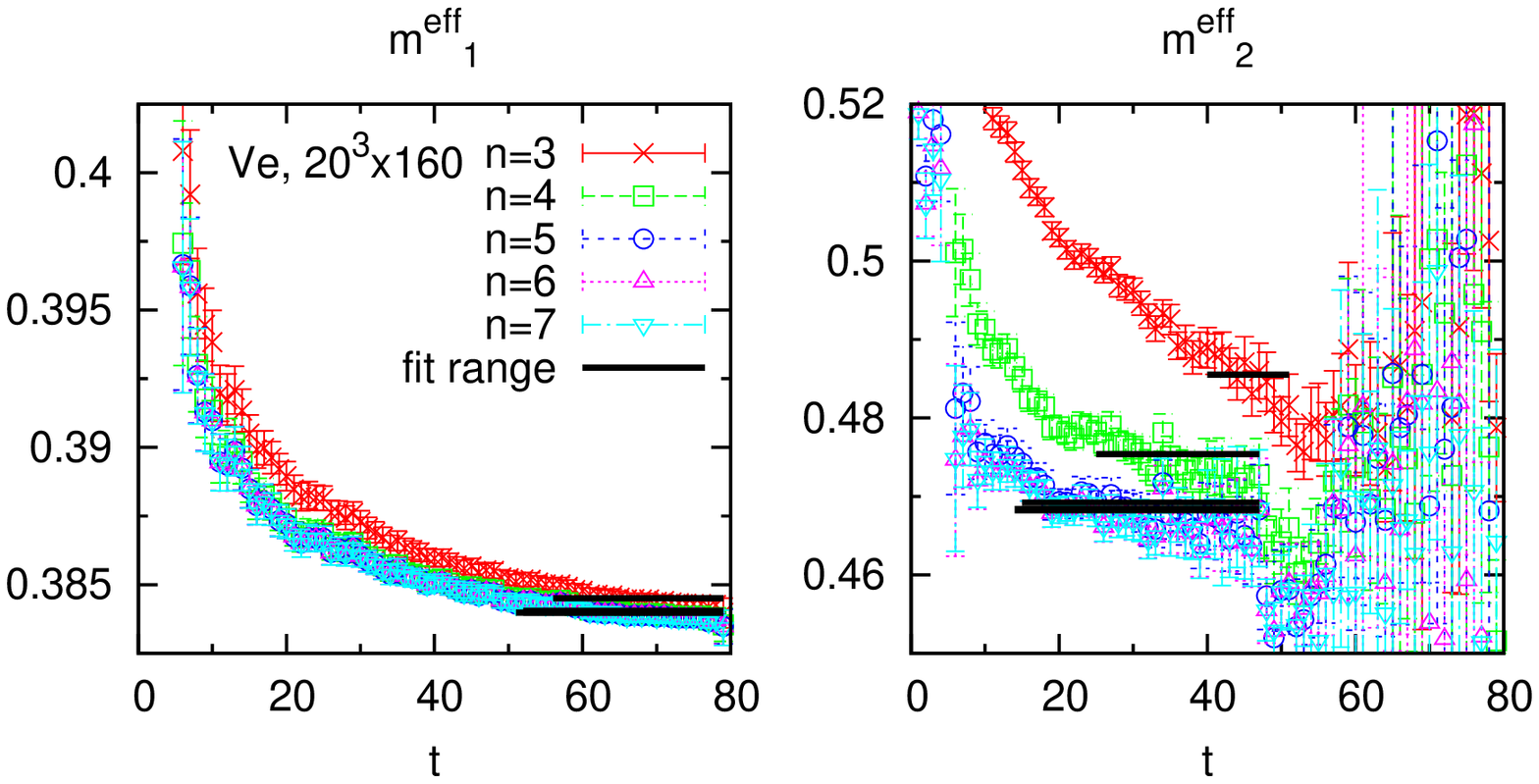}
  \includegraphics[width=44mm, angle=-90]{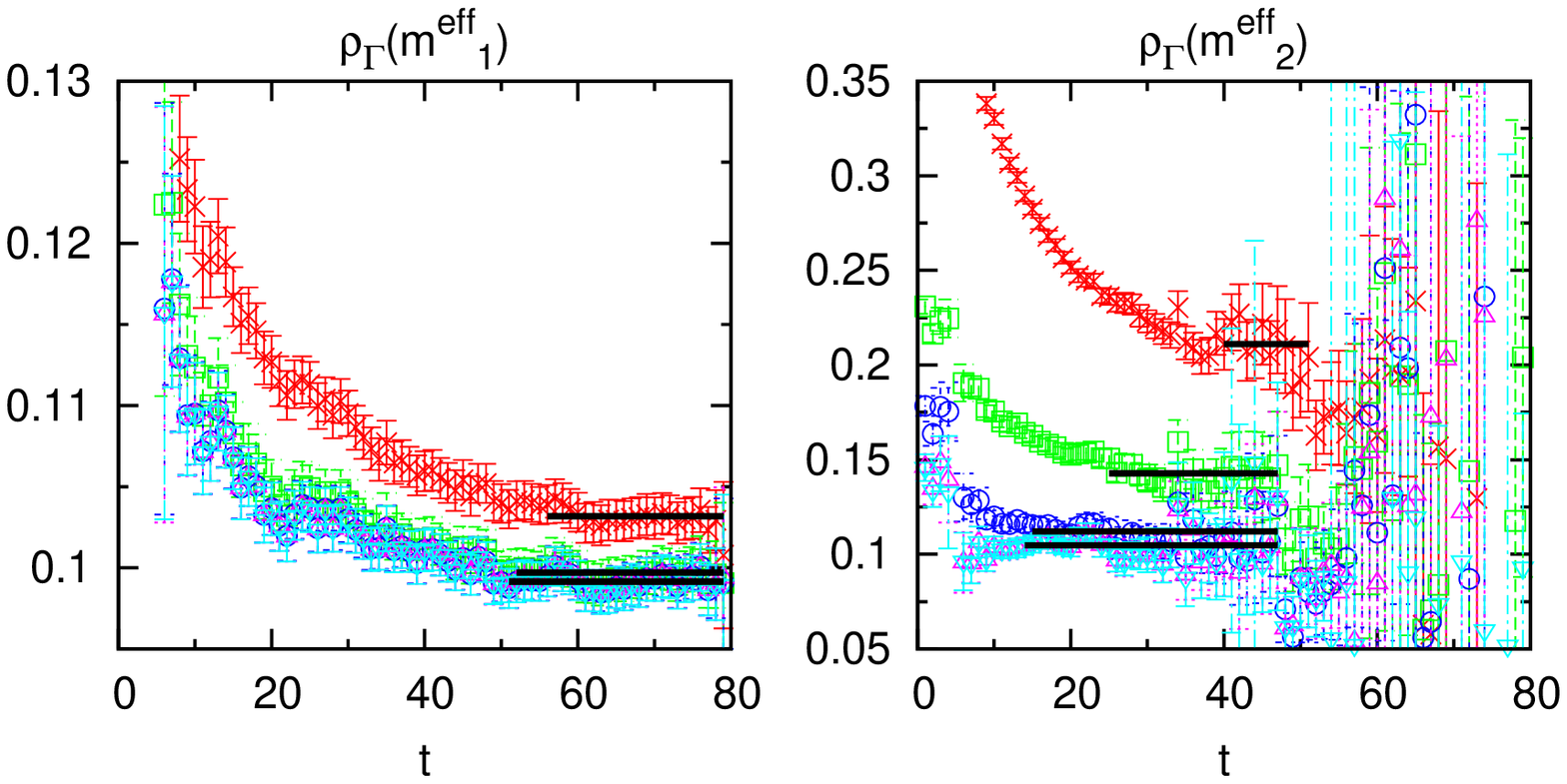}
  \caption{The same as FIG. \ref{fit_Ps} for the Ve channel.} 
  \label{fit_Ve}
 \end{center}
\end{figure} 
\begin{figure}[tbh]
 \begin{center}
  \includegraphics[width=44mm, angle=-90]{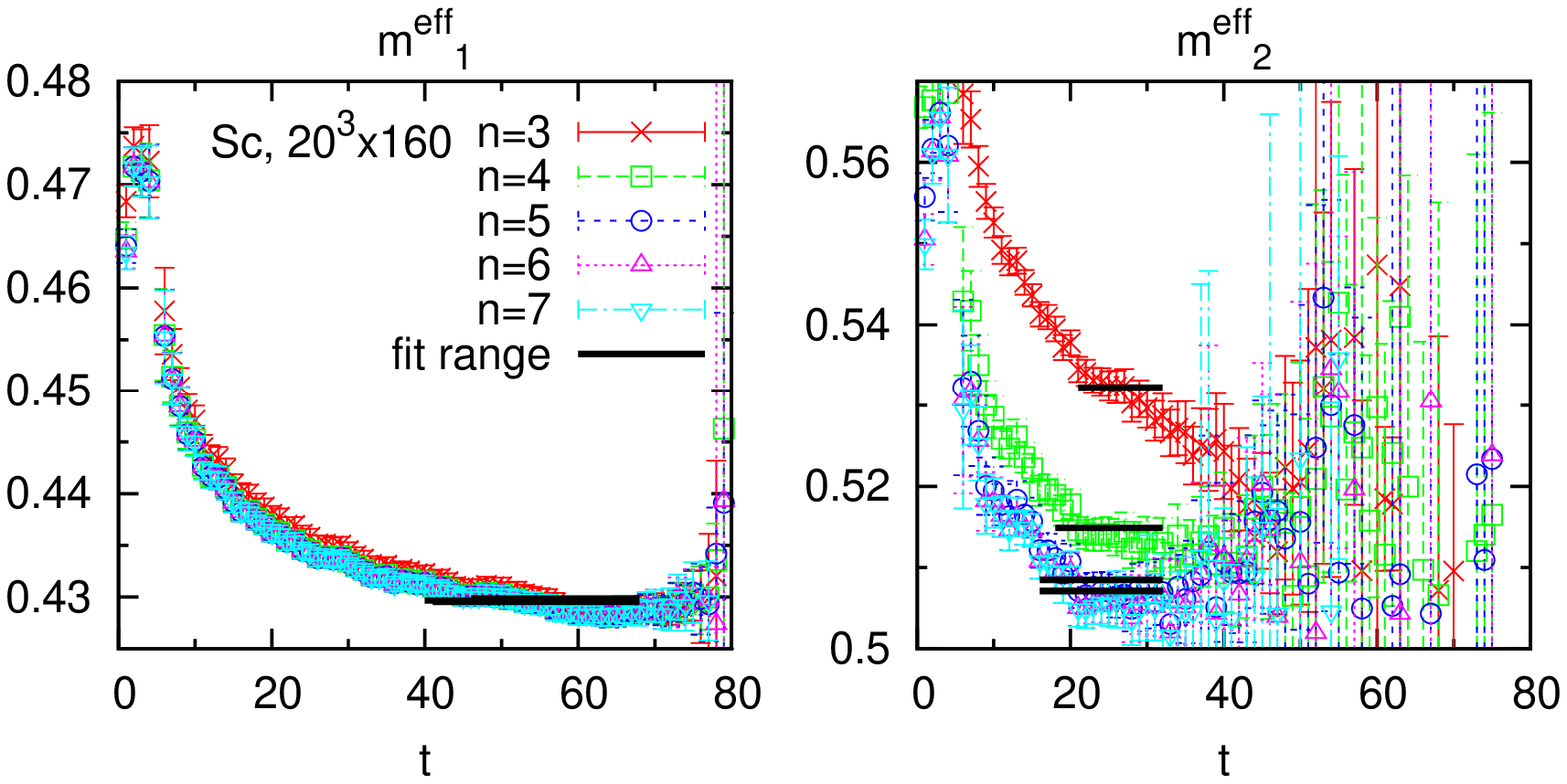}
  \includegraphics[width=44mm, angle=-90]{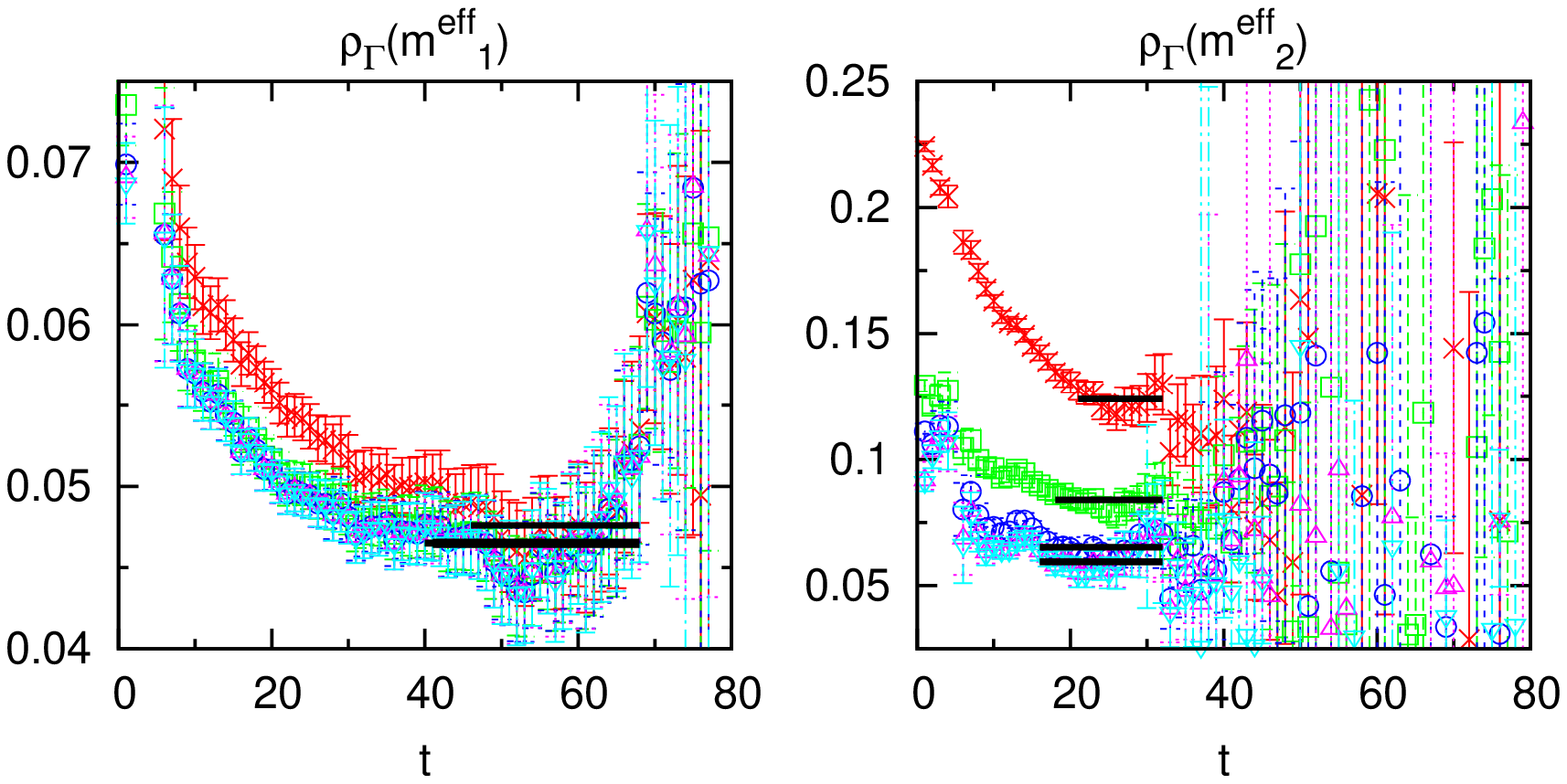}
  \caption{The same as FIG. \ref{fit_Ps} for the Sc channel.} 
  \label{fit_Sc}
 \end{center}
\end{figure} 
\begin{figure}[tbh]
 \begin{center}
  \includegraphics[width=44mm, angle=-90]{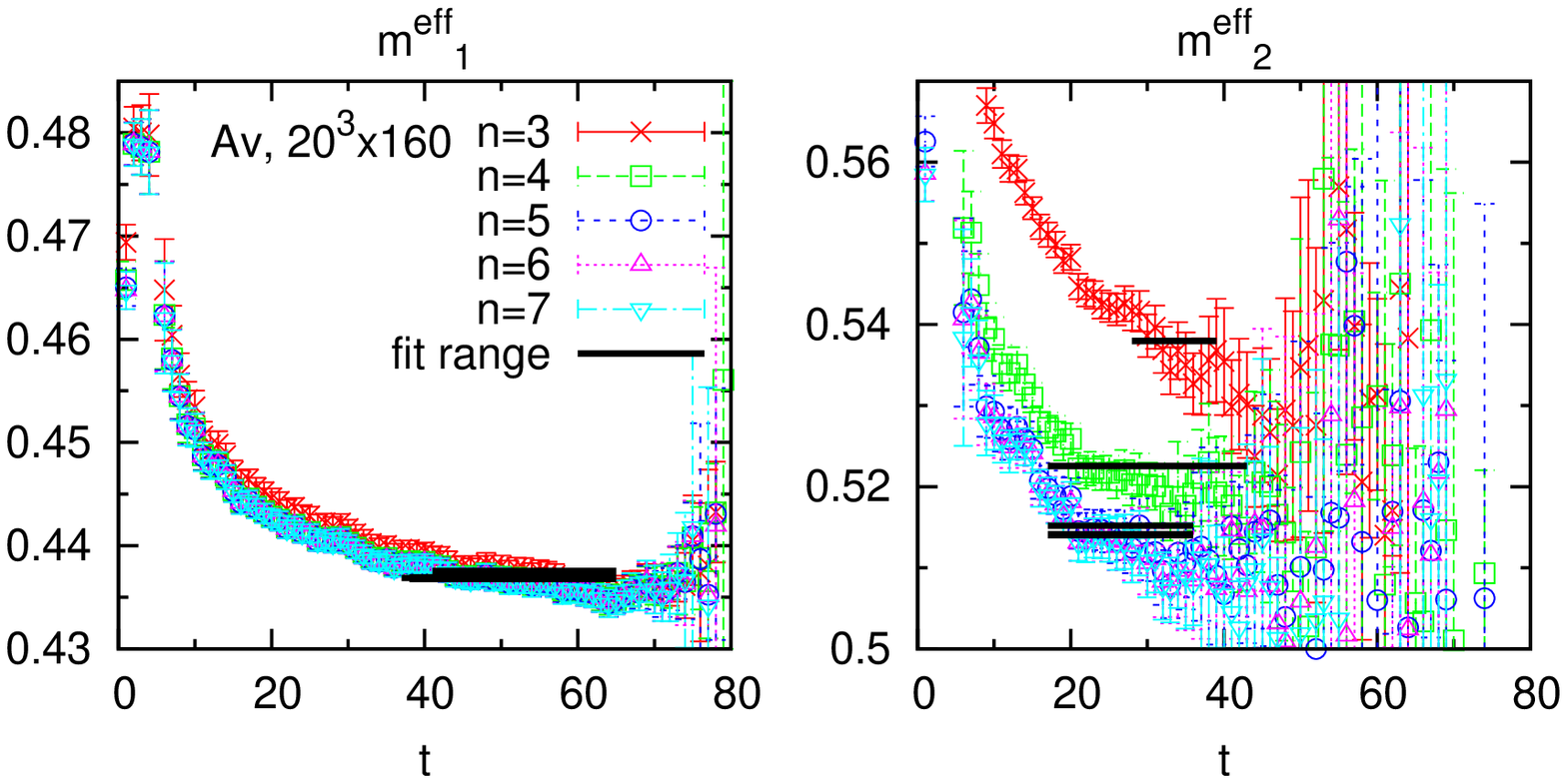}
  \includegraphics[width=44mm, angle=-90]{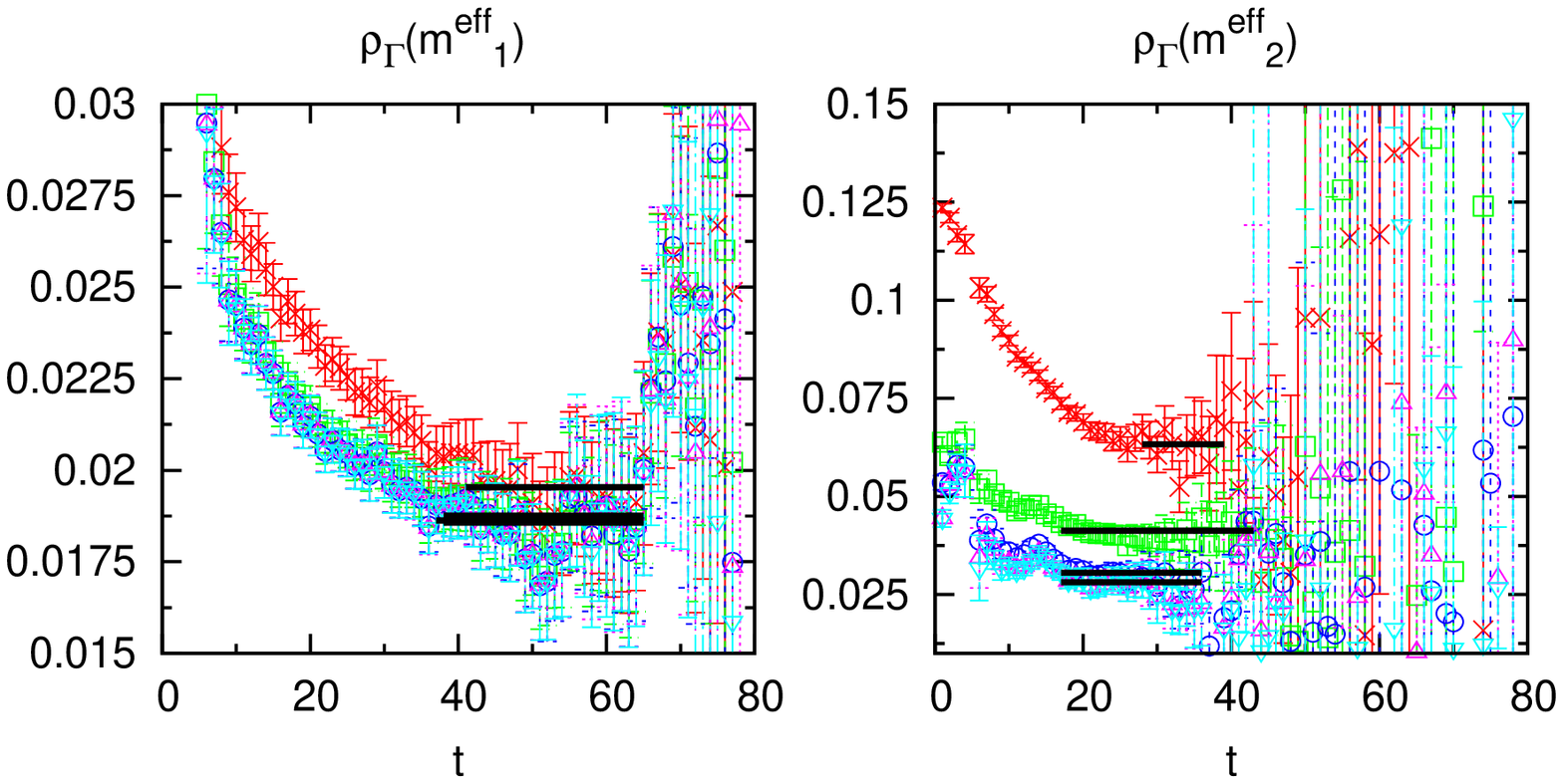}
  \caption{The same as FIG. \ref{fit_Ps} for the Av channel.} 
  \label{fit_Av}
 \end{center}
\end{figure} 

\begin{figure}[tbh]
 \begin{center}
  \includegraphics[width=62mm, angle=-90]{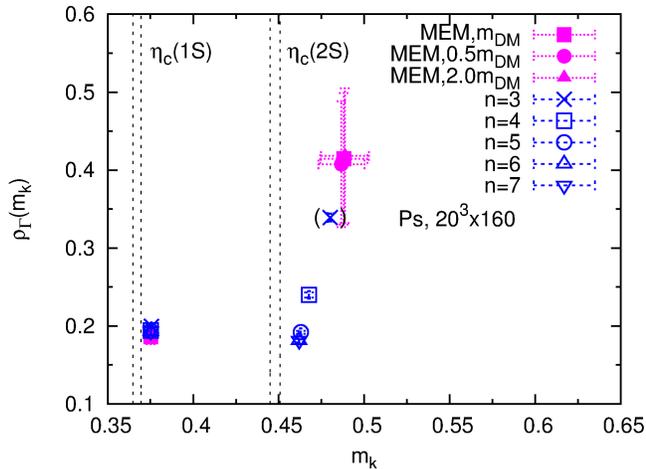}
  \vspace{-3mm}
  \caption{Charmonium spectral function at the ground and the first excited states for Ps channel at zero temperature obtained on the $20^3\times 160$ lattice.
  The reference point is at $t_0=5$. Cross, square, circle, triangle, and downward triangle symbols indicate the data by the variational method with $n=3,4,5,6,7$, respectively,
  and solid symbols indicate the MEM results. 
  The vertical dashed lines indicate the range of experimental masses for the $\eta_c({\rm 1S})$ and $\eta_c({\rm 2S})$ mesons.} 
  \label{var_mem_Ps}
 \end{center}
\end{figure}
\begin{figure}[tbh]
 \begin{center}
  \includegraphics[width=62mm, angle=-90]{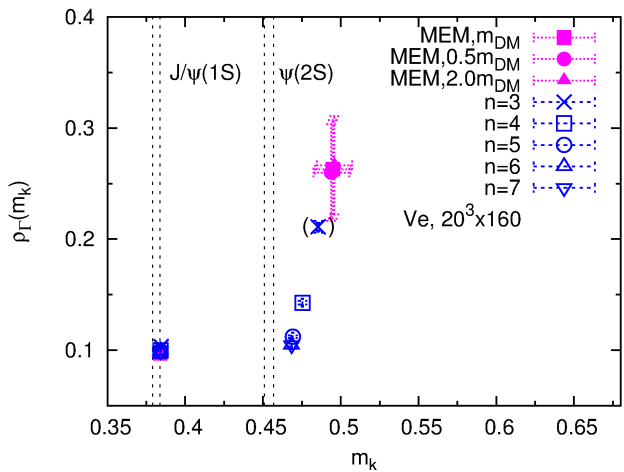}
  \vspace{-3mm}
  \caption{The same as Fig.~\ref{var_mem_Ps} for the Ve channel.
  The vertical dashed lines indicate the range of experimental masses for the $J/\psi({\rm 1S})$ and $\psi({\rm 2S})$ mesons.} 
  \label{var_mem_Ve}
 \end{center}
\end{figure} 
\begin{figure}[tbh]
 \begin{center}
  \includegraphics[width=62mm, angle=-90]{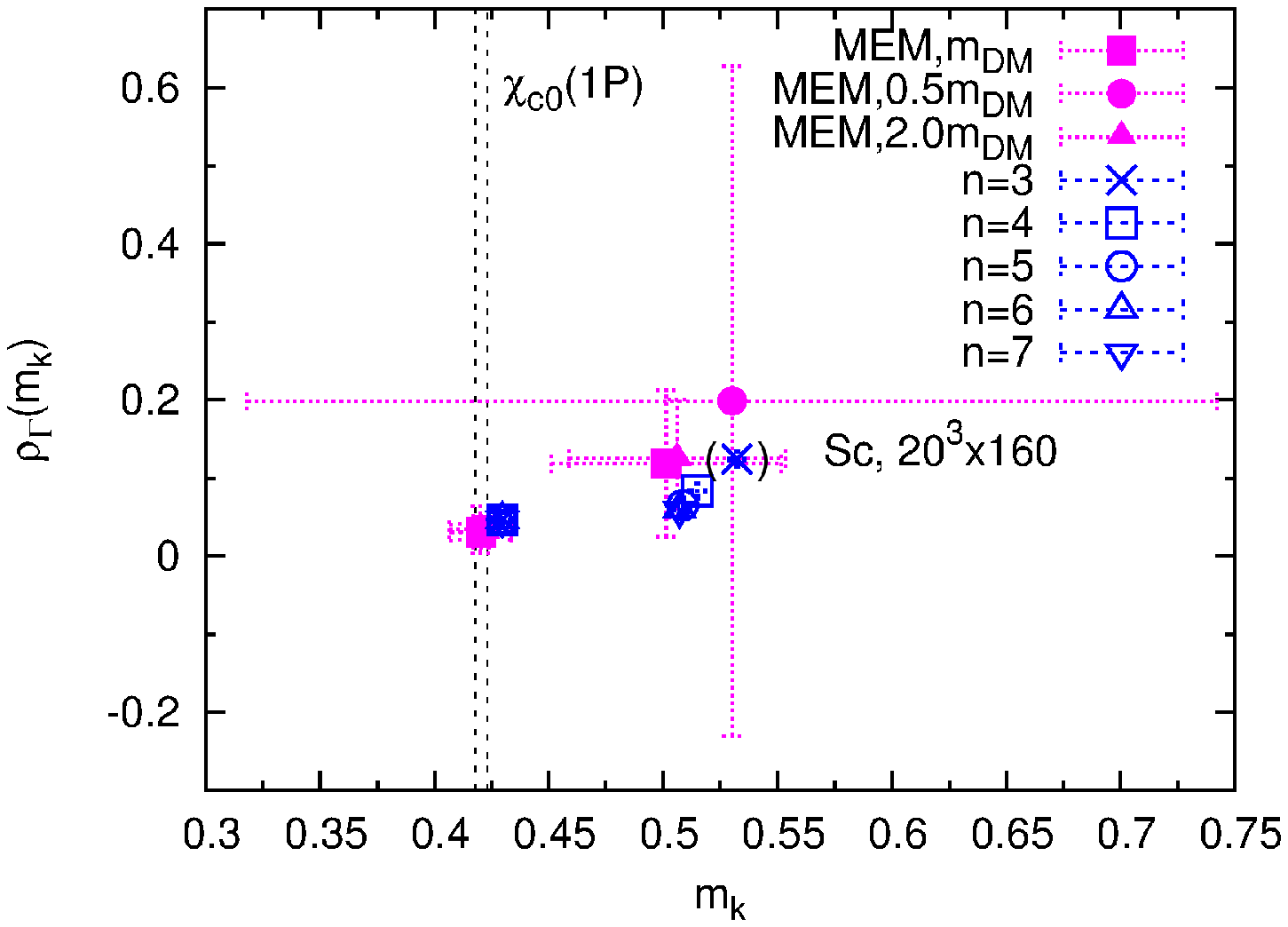}
  \vspace{-3mm}
  \caption{The same as Fig.~\ref{var_mem_Ps} for the Sc channel.
  The vertical dashed lines indicate the range of experimental mass for the $\chi_{c0}({\rm 1P})$ meson.} 
  \label{var_mem_Sc}
 \end{center}
\end{figure} 
\begin{figure}[tbh]
 \begin{center}
  \includegraphics[width=62mm, angle=-90]{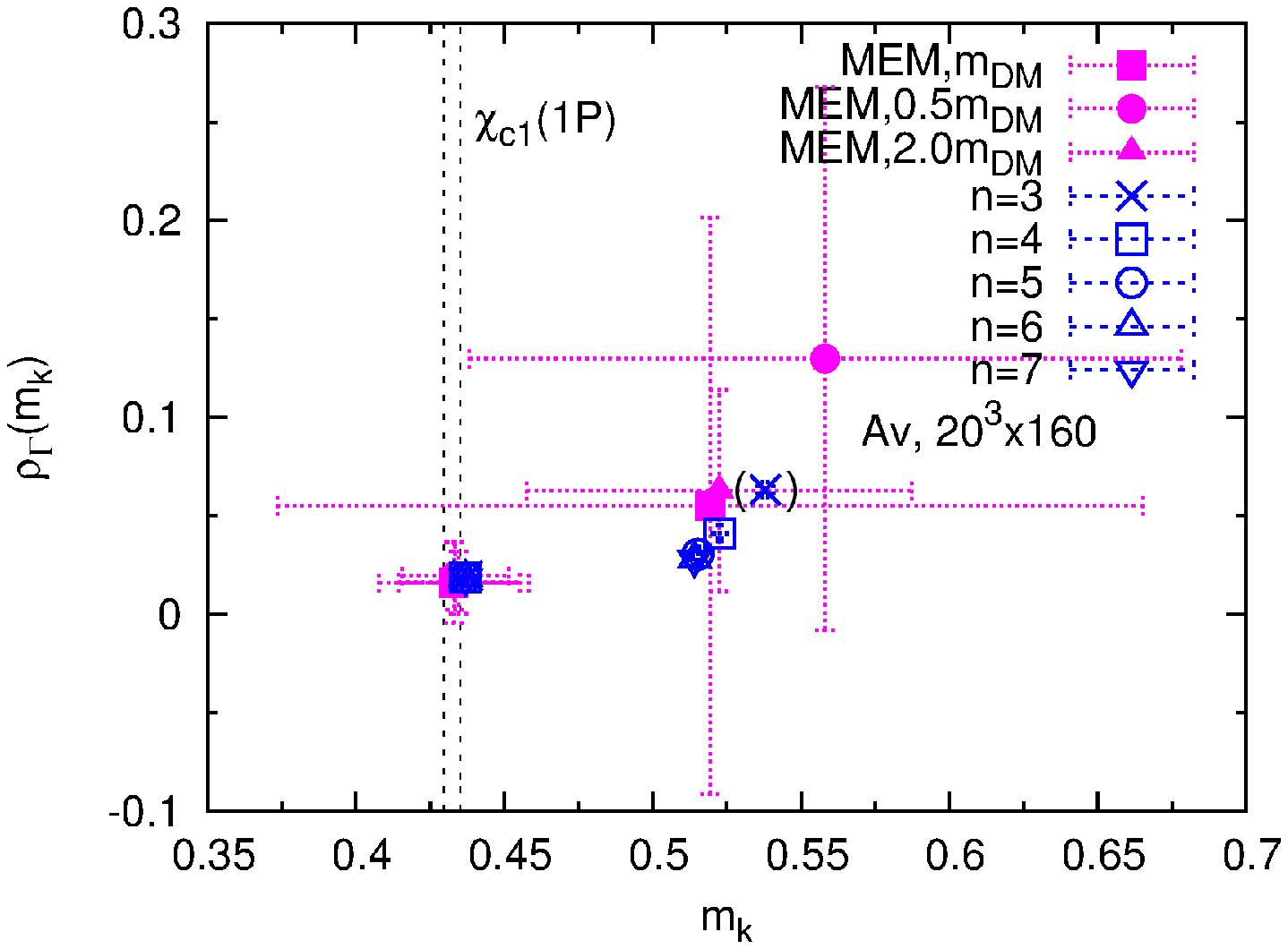}
  \vspace{-3mm}
  \caption{The same as Fig.~\ref{var_mem_Ps} for the Av channel.
  The vertical dashed lines indicate the range of experimental mass for the $\chi_{c1}({\rm 1P})$ meson.} 
  \label{var_mem_Av}
 \end{center}
\end{figure} 

\section{\label{sec:results}Charmonium spectral functions}

In this section, we study charmonium spectral functions in quenched QCD and compare the results of the variational method with those obtained by the conventional MEM.

\subsection{\label{sec:setup}Simulation parameters}

We perform simulations on anisotropic $20^3\times N_t$ lattices with the renormalized anisotropy $\xi=4$ adopting the standard plaquette gauge action.
We study at $\beta=6.10$ where the spatial lattice spacing determined by the Sommer scale $r_0 = 0.5$ fm \cite{Sommer} is $a_s=0.0970(5)$ fm ($a^{-1}_s=2.030(13)$ GeV).
Our spatial volume is thus about $(2\; \mathrm{fm})^3$.
The bare anisotropy for $\xi=4$ is $\gamma_G=3.2108$ \cite{action}.
For valence quarks, we adopt an $O(a)$-improved Wilson quark action. 
We set the bare fermionic anisotropy $\gamma_F=4.94$
and tree-level tadpole improved clover coefficients $c_E=3.164$ and $c_B=1.911$ to realize $\xi=4$
(see Ref. \cite{action} for the definitions of the coupling parameters).
In this study, we set the Wilson parameter $r=1$ to suppress lattice artifacts in excited charmonia \cite{Karsch}.
We study at $\kappa=0.10109$, which corresponds to the physical charm quark mass on an isotropic lattice with a similar spatial lattice spacing.

For the temporal lattice size, 
we adopt $N_t=160$ for the zero-temperature simulation, and $N_t = 32$, 26, and 20 for finite-temperature simulations at $T\approx 0.88 T_c$, $1.1 T_c$, and $1.4 T_c$, respectively.
Here the critical temperature $T_c$ is determined by the peak position of Polyakov loop susceptibility and corresponds to $N_t \approx 28$. 
After 20 000 sweeps for thermalization, we generate 299 configurations at zero temperature and 800 configurations at finite temperatures separated by 500 pseudo-heat-bath sweeps.
Our simulation parameters are summarized in Table~\ref{sim_param}.
To calculate smeared operators, we use the Coulomb gauge.
Statistical errors for spectral functions are estimated by a jackknife method.

\subsection{\label{sec:zero_temp}Zero temperature}

At zero temperature, we calculate the locations and the heights of the peaks for charmonia spectral functions up to the first excited state for Ps, Ve, Sc and Av channels.
We first calculate the spectral functions with the conventional MEM \cite{MEM} using meson correlation functions for point operators.
We use the range $t=1$--60 for Ps and Ve channels, and $t=3$--60 for Sc and Av channels,
because the latter correlation functions suffer from lattice artifacts at $t\sim1$.
For the default model $m(\omega)$, we adopt
\begin{equation}
m(\omega) = \alpha m_{\rm DM}\, \omega^2,
\label{eq:DM}
\end{equation}
where $m_{DM} = 4.2$ for Ps and Sc channels and 2.4 for Ve and Av channels, as determined by the asymptotic behavior of meson correlation functions in the perturbation theory at $\alpha=1$ \cite{lqcd1, MEM}.
We estimate statistical errors by the jackknife method.
To estimate systematic errors due to the choice of the default model, we multiply the factor $\alpha$ and vary it in the range 0.5--2.0.
We have checked that when the default model is fixed, the results are stable under variations of various parameters in the MEM.

Our results of spectral functions by MEM for Ps, Ve, Sc and Av channels are shown in Figs.~\ref{mem_Ps} and \ref{mem_Sc} for the cases $\alpha=1$, 0.5 and 2.0. 
We find that for the S-waves, the peaks are well isolated  up to the first excited states and the results are approximately stable.
But, for P-waves, the peaks are not isolated and peaks for excited states are not stable.
We identify $m_k$ with the peak position defined by the maxima of the spectral function, 
and $\rho_{\Gamma}(m_k)$ with the area of the peak at $m_k$. 
For the P-waves for which the peaks are not well isolated, we divide the spectral functions into each peak at the minima to compute the area. 

To calculate spectral functions with the variational method, we adopt the Gaussian smearing function (\ref{eq:Gsmearing}) with the smearing parameters listed in Table~\ref{smearing_param}.
The effective masses and effective spectral functions with $t_0=5$ are shown in Figs.~\ref{fit_Ps}--\ref{fit_Av}.
To extract the plateau, we fit the data in the range shown by horizontal solid lines in Figs.~\ref{fit_Ps}--\ref{fit_Av} for each charmonium.
Here, we choose the same fit ranges for effective masses and corresponding effective spectral functions.
The fit ranges [$t_{\mathrm{min}}$,$t_{\mathrm{max}}$] are chosen as follows. We first determine $t_{\mathrm{max}}$ as the upper bound of
the plateau of $m^{\mathrm{eff}}_k$ and $\rho_{\Gamma}(m^{\mathrm{eff}}_k)$ for each $k$. 
Then, we study $t_{\mathrm{min}}$-dependence of $m_k$ and $\rho_{\Gamma}(m_k)$. As expected from Figs.11-14, when $n$ is small,
we sometimes observe that $m_k$ keeps decreasing with increasing $t_{\mathrm{min}}$ up to $t_{\mathrm{max}}$. When $n$ is sufficiently large, however,
$m_k$ and $\rho_{\Gamma}(m_k)$ are stable in a wide range in the sense that $t_{\mathrm{min}}$-dependences are smaller than, or at least comparable with,
the small statistical errors, though the statistical errors become large when $t_{\mathrm{min}}$ becomes close to $t_{\mathrm{max}}$.
Because a simple criterion to define a plateau region is not available, in this study, we just choose $t_\mathrm{min}$, where $\chi^2/{\rm dof}$ for $m_k$
becomes closest to 1, and reject the data by adding brackets in the plots when a systematic tendency to deviate from the plateau is observed beyond statistical
errors by increasing $t_{\mathrm{min}}$ towards $t_{\mathrm{max}}$. With the present lattice size and statistical accuracy, we cannot exclude a mild
decrease of $m_k$ and $\rho_{\Gamma}(m_k)$ by choosing the fit range at larger $t$. Therefore, our $m_k$ and $\rho_{\Gamma}(m_k)$ should be regarded as upper
bounds for them. Resulting ranges of the value of $\chi^2/{\rm dof}$ are 0.82--1.1 and 0.045--0.54 for $m_k$ and $\rho_{\Gamma}(m_k)$, respectively.
We confirm that the variations of $m_k$ and $\rho_{\Gamma}(m_k)$ under a change of $t_{\mathrm{min}}$ to $t_{\mathrm{min}}\pm1$ are less than 0.2\% and 2\%, respectively.

Our results for charmonium spectral functions are summarized in Figs.~\ref{var_mem_Ps}--\ref{var_mem_Av}. 
The errors for the variational method's data are smaller than the symbols.
The symbols in the brackets are not asymptotic.
For the results from MEM, statistical errors for $m_k$ and $\rho_\Gamma(m_k)$ are given for each values of $\alpha$. 
We note that the statistical errors in $m_k$ and $\rho_\Gamma(m_k)$ have strong positive correlation, i.e., the errors actually shape a thin oval inclined rightwards.
The experimental values of corresponding charmonium masses with their errors are shown by vertical dashed lines \cite{PDG2010}.
In this study, the charm quark mass is adjusted to approximately reproduce the experimental J/$\psi$ mass.
On the other hand, masses of $\eta_c$, etc., show slight deviations from experiment, in accordance with 
the previous observation with $O(a)$-improved Wilson quarks that
the charmonium hyperfine splitting is smaller than experiment in quenched QCD  \cite{charm_hfs}.

We find that the results for the ground states are well consistent with each other between the MEM and the variational method with all $n$ studied. 
On the other hand, for the first excited states, the variational method leads to results discrepant from those of MEM.
For the S-waves, we find that the discrepancy is beyond the errors estimated with the MEM and becomes larger with increasing $n$.
We find that the results of the variational method converge to a point close to the experimental values.
This suggests that the inclusion of higher states in the variational approach helps improve the signals for excited states.
We also note that the results of MEM approximately corresponds to those of the variational method in the limit of small $n$, in accordance with the fact that the studies with MEM are based on the information of point-source correlation functions only.
The results for P-wave excited states are similar, but the errors in the MEM results are quite large to draw a definite conclusion.

\begin{figure}[tbh]
 \begin{center}
  \includegraphics[width=57mm, angle=-90]{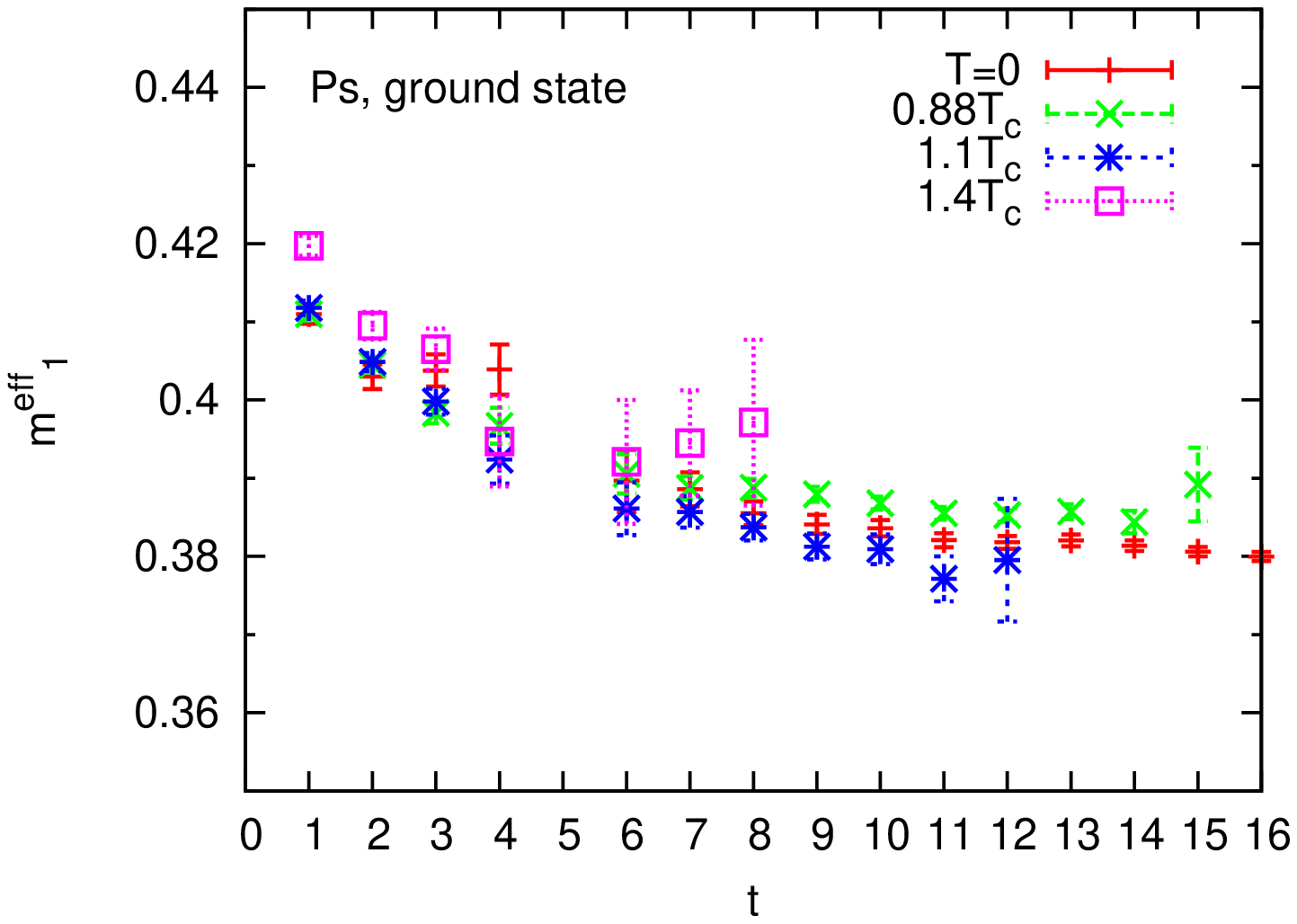}
  \includegraphics[width=57mm, angle=-90]{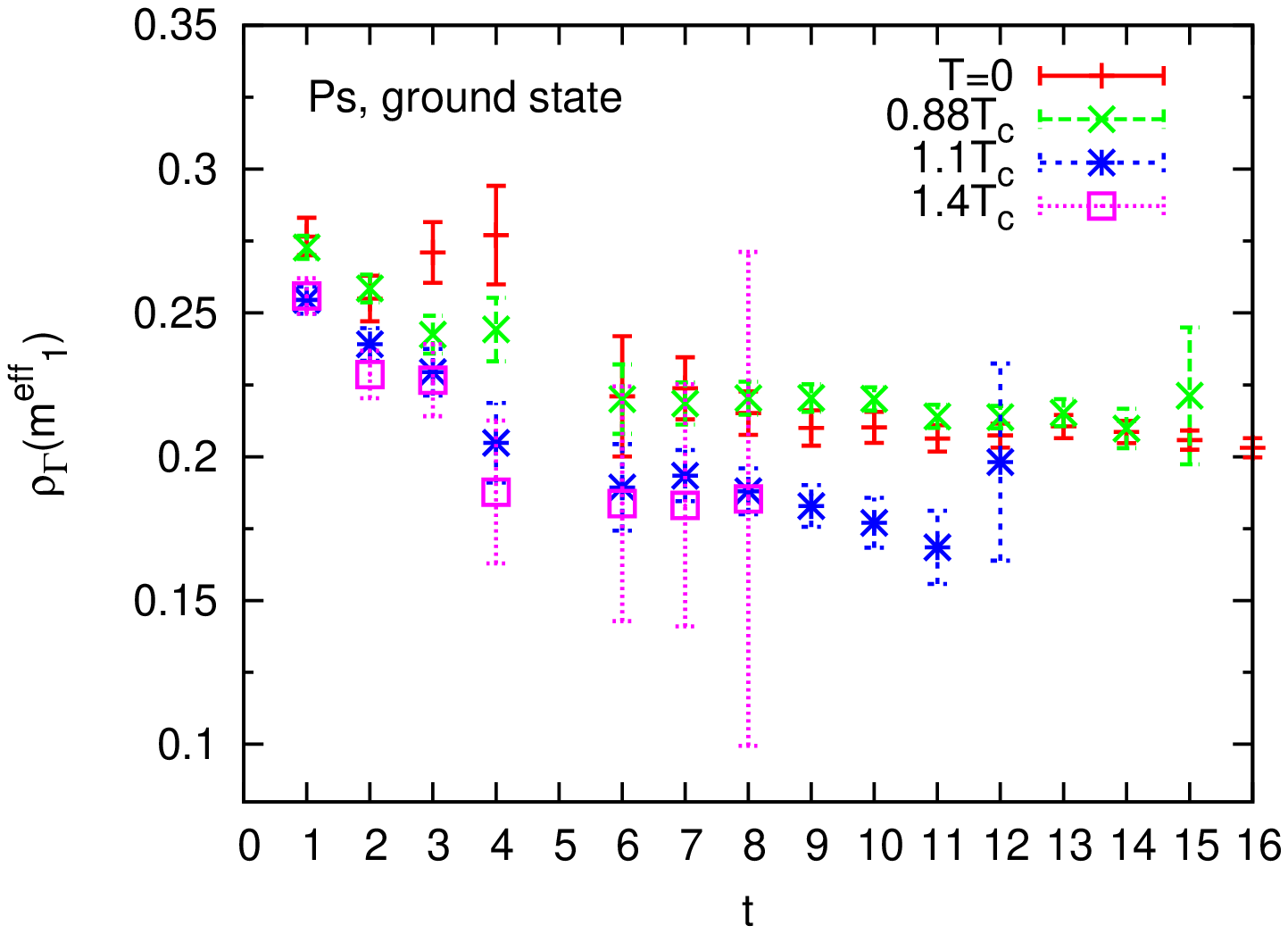}
  \caption{$m^{\mathrm{eff}}_1(t,t_0)$ and $\rho_{\Gamma}(m^{\mathrm{eff}}_1(t,t_0))$ for the Ps channel with $n=7$ and $t_0=5$. The plus, cross, asterisk, and square symbols indicate
  the data at $T=0$, $0.88T_c$, $1.1T_c$, and $1.4T_c$, respectively.} 
  \label{temp_mass_Ps}
 \end{center}
\end{figure}
\begin{figure}[tbh]
 \begin{center}
  \includegraphics[width=57mm, angle=-90]{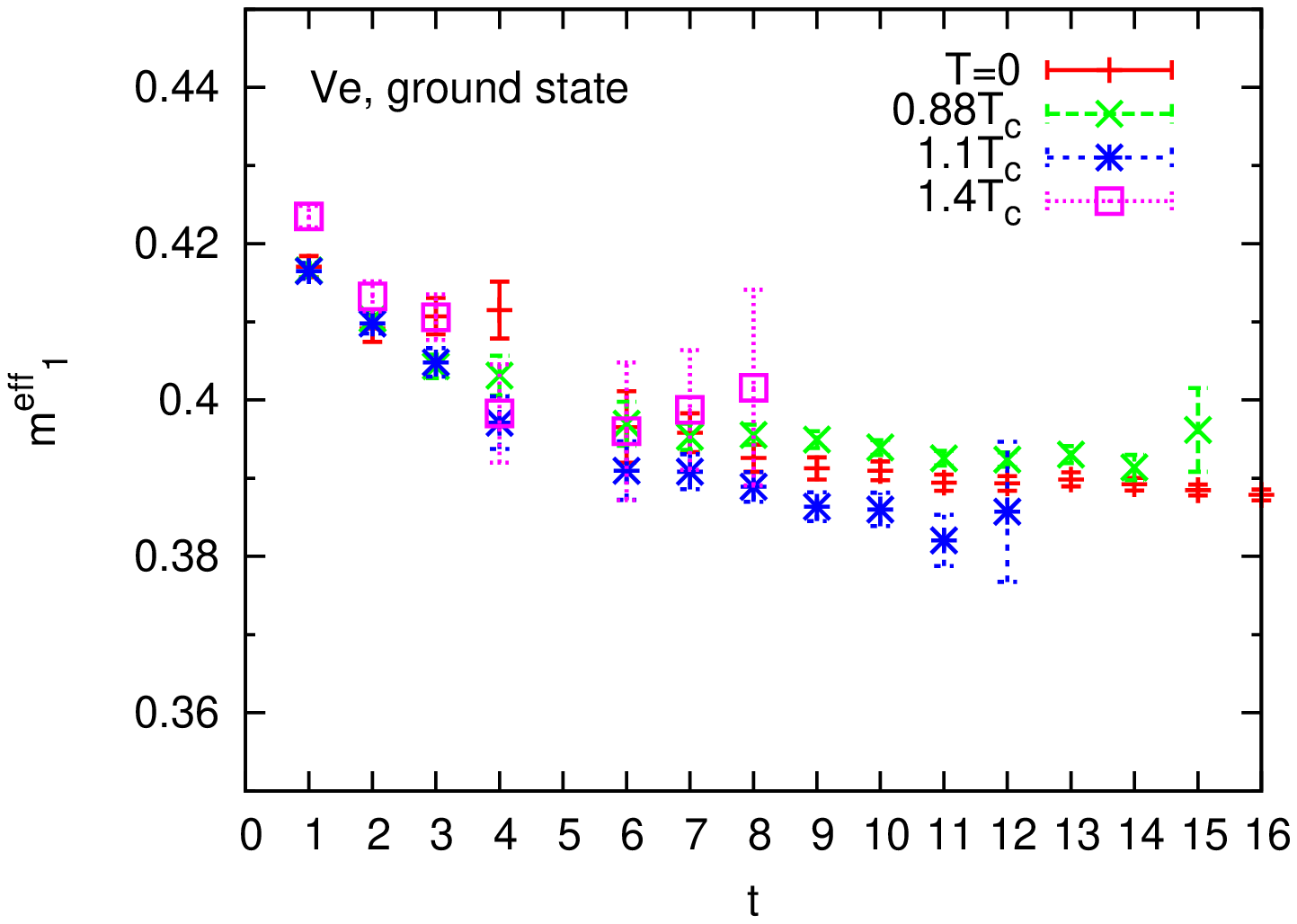}
  \includegraphics[width=57mm, angle=-90]{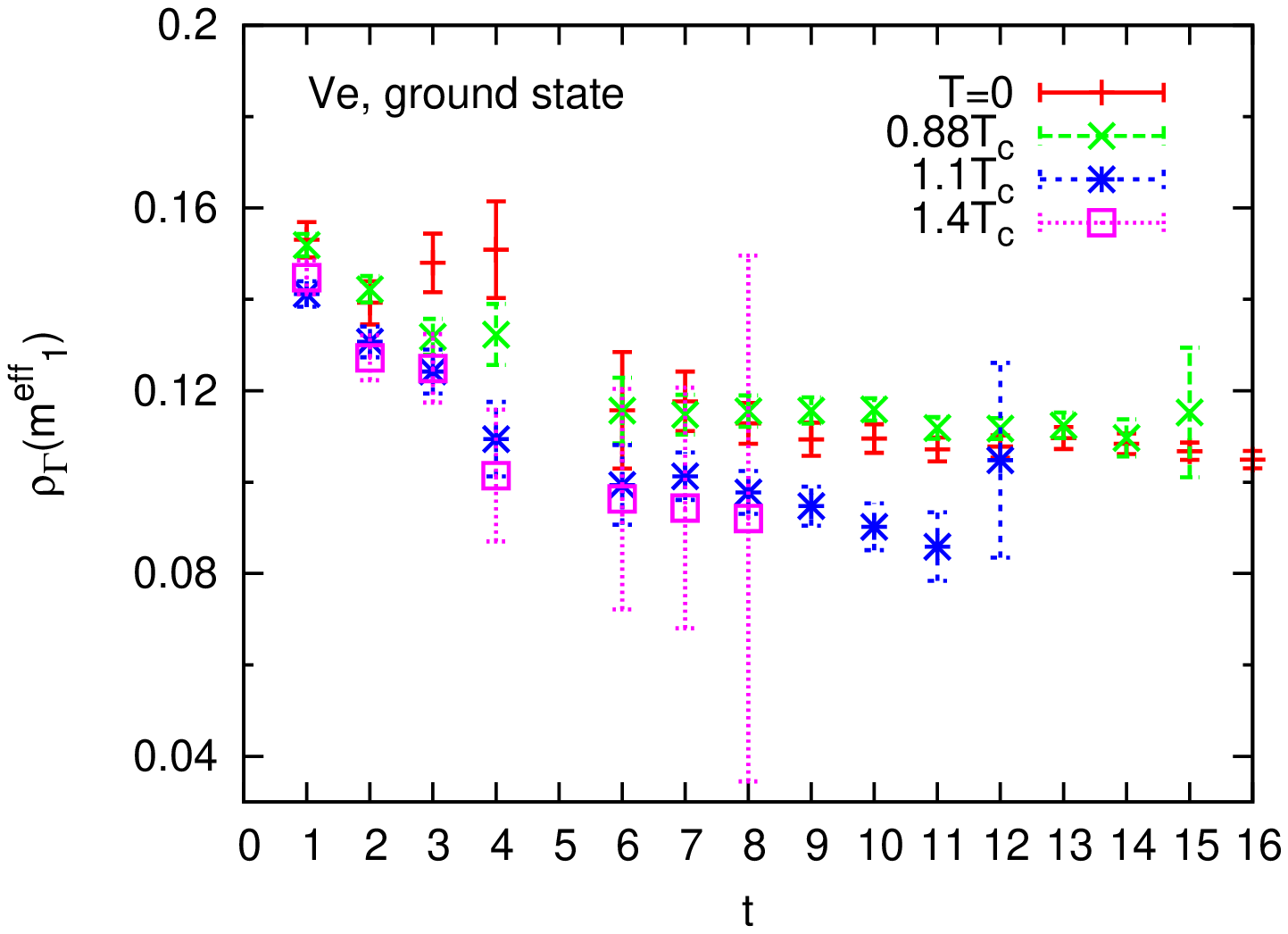}
  \caption{The same as Fig.~\ref{temp_mass_Ps} for the Ve channel.} 
  \label{temp_mass_Ve}
 \end{center}
\end{figure} 

\begin{figure}[tbh]
 \begin{center}
  \includegraphics[width=55mm, angle=-90]{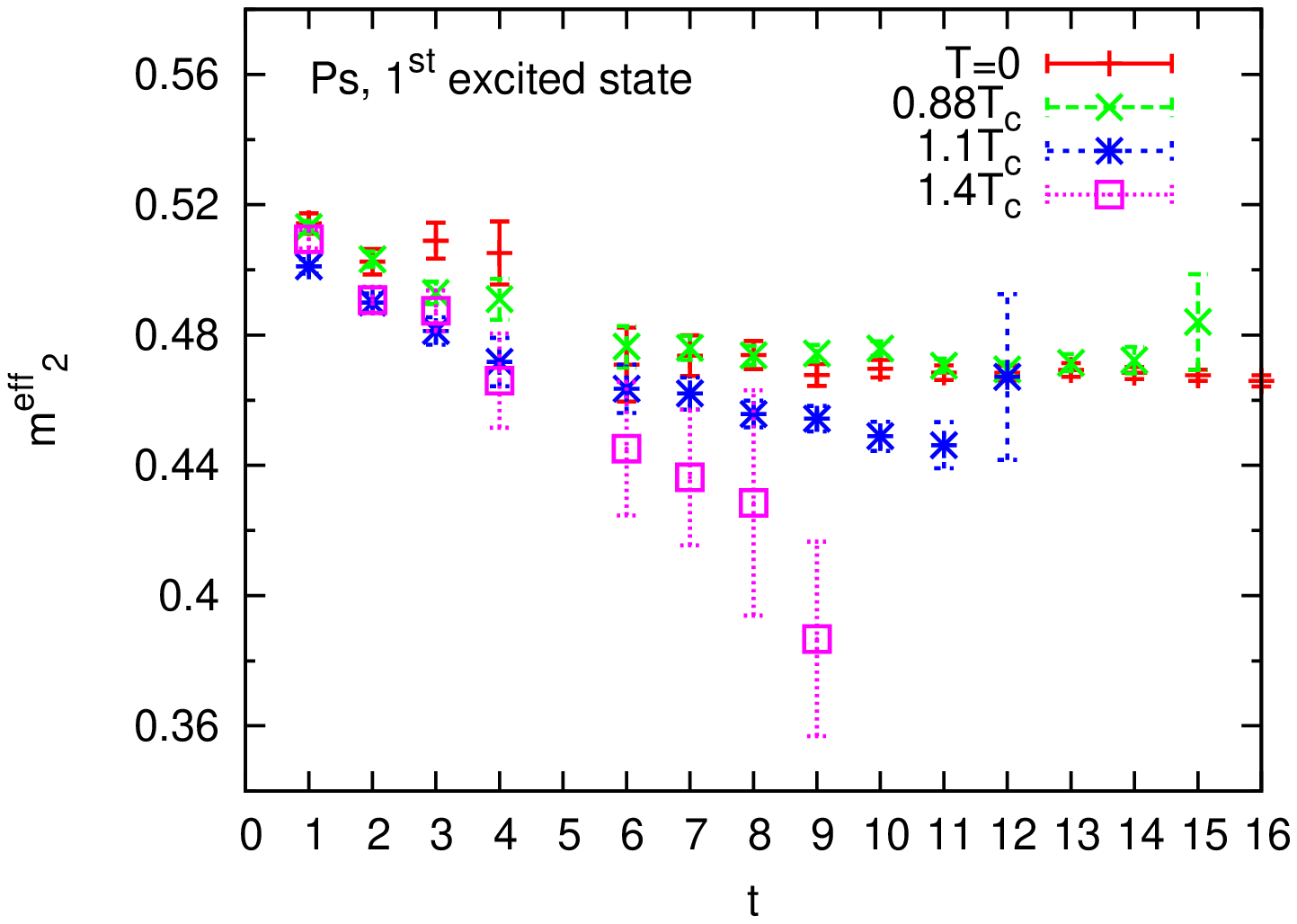}
  \includegraphics[width=55mm, angle=-90]{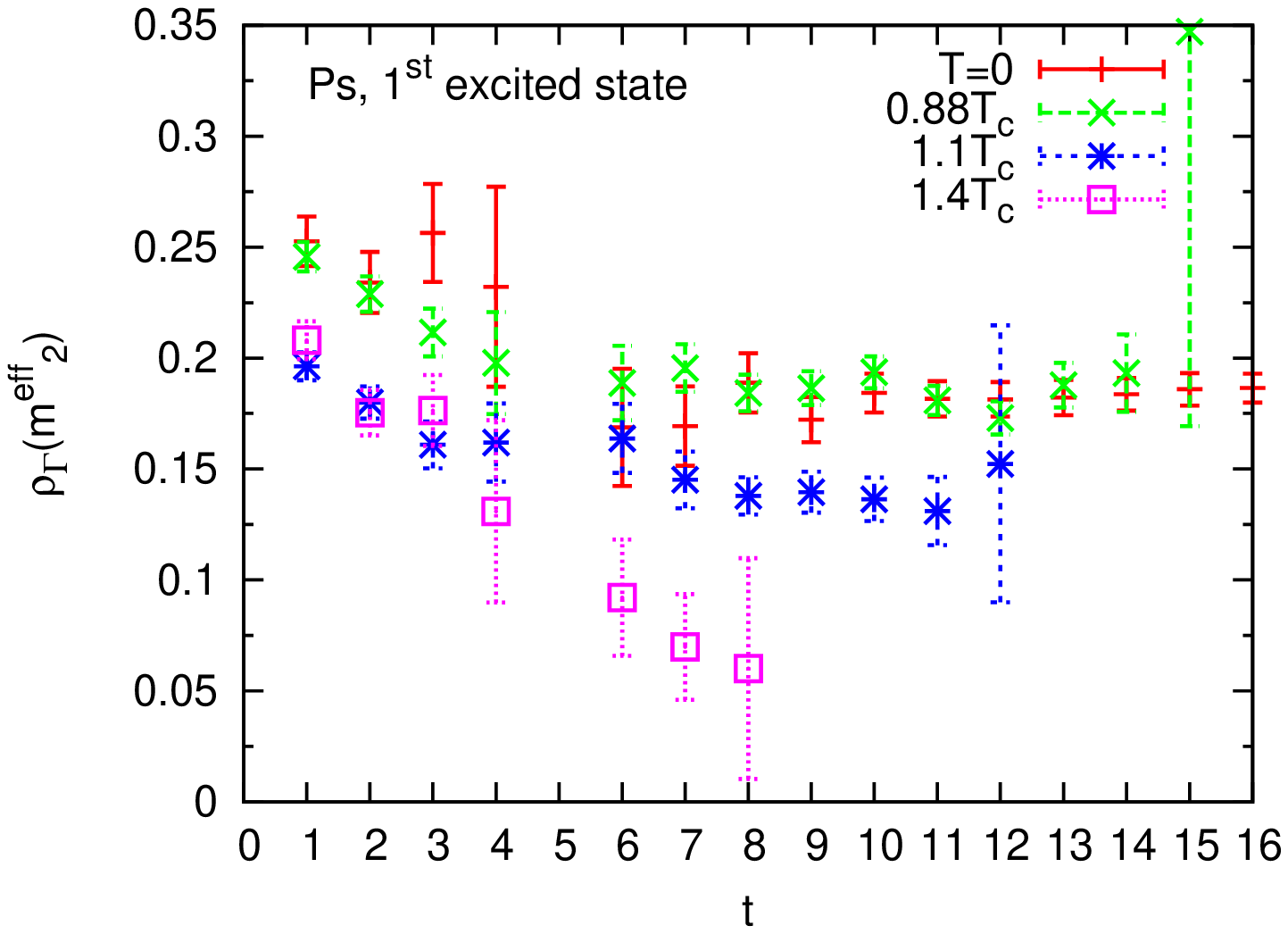}
  \caption{$m^{\mathrm{eff}}_2(t,t_0)$ and $\rho_{\Gamma}(m^{\mathrm{eff}}_2(t,t_0))$ for the Ps channel.} 
  \label{temp_mass_Ps1}
 \end{center}
\end{figure}

\begin{figure}[tbh]
 \begin{center}
  \includegraphics[width=55mm, angle=-90]{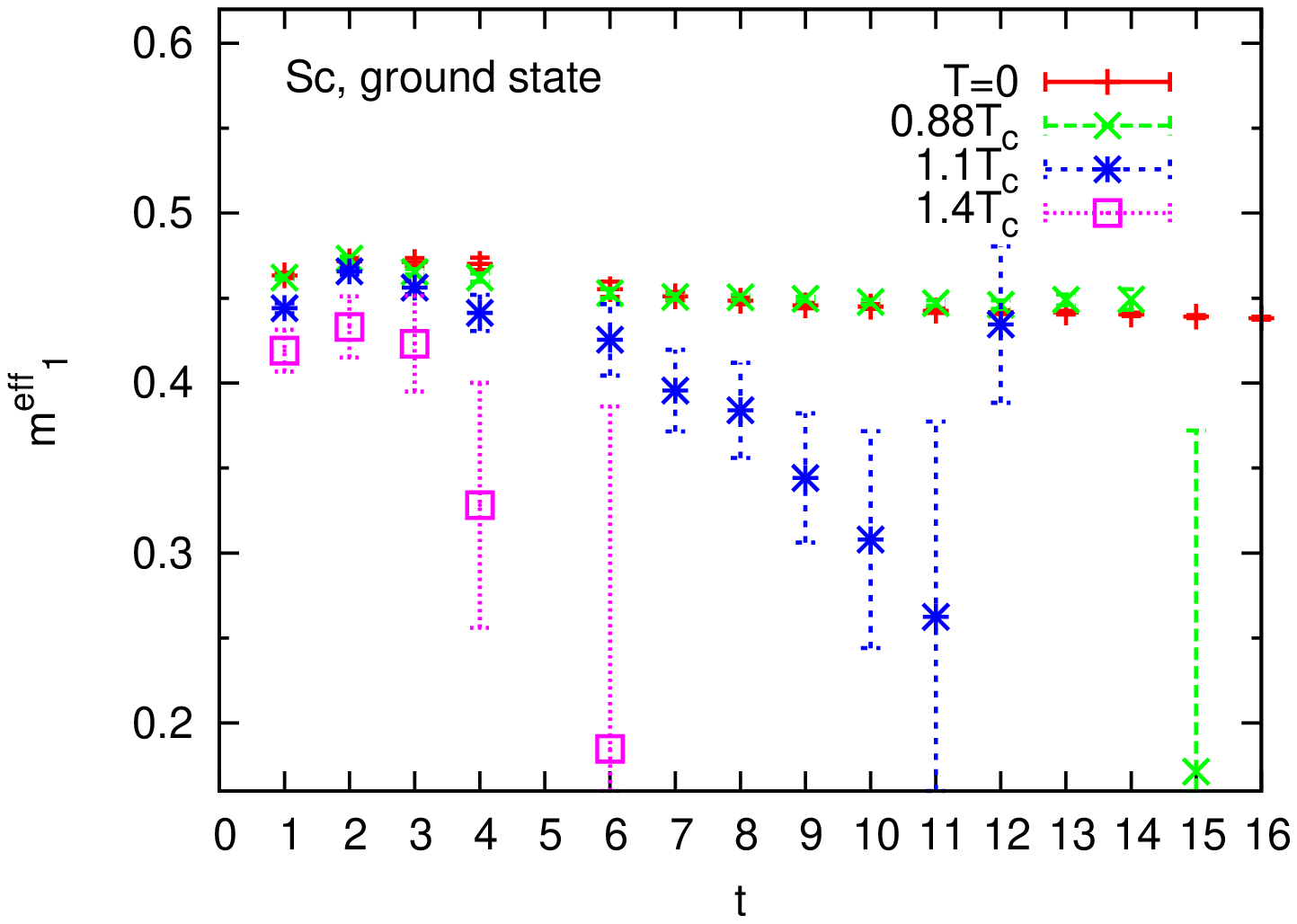}
  \includegraphics[width=55mm, angle=-90]{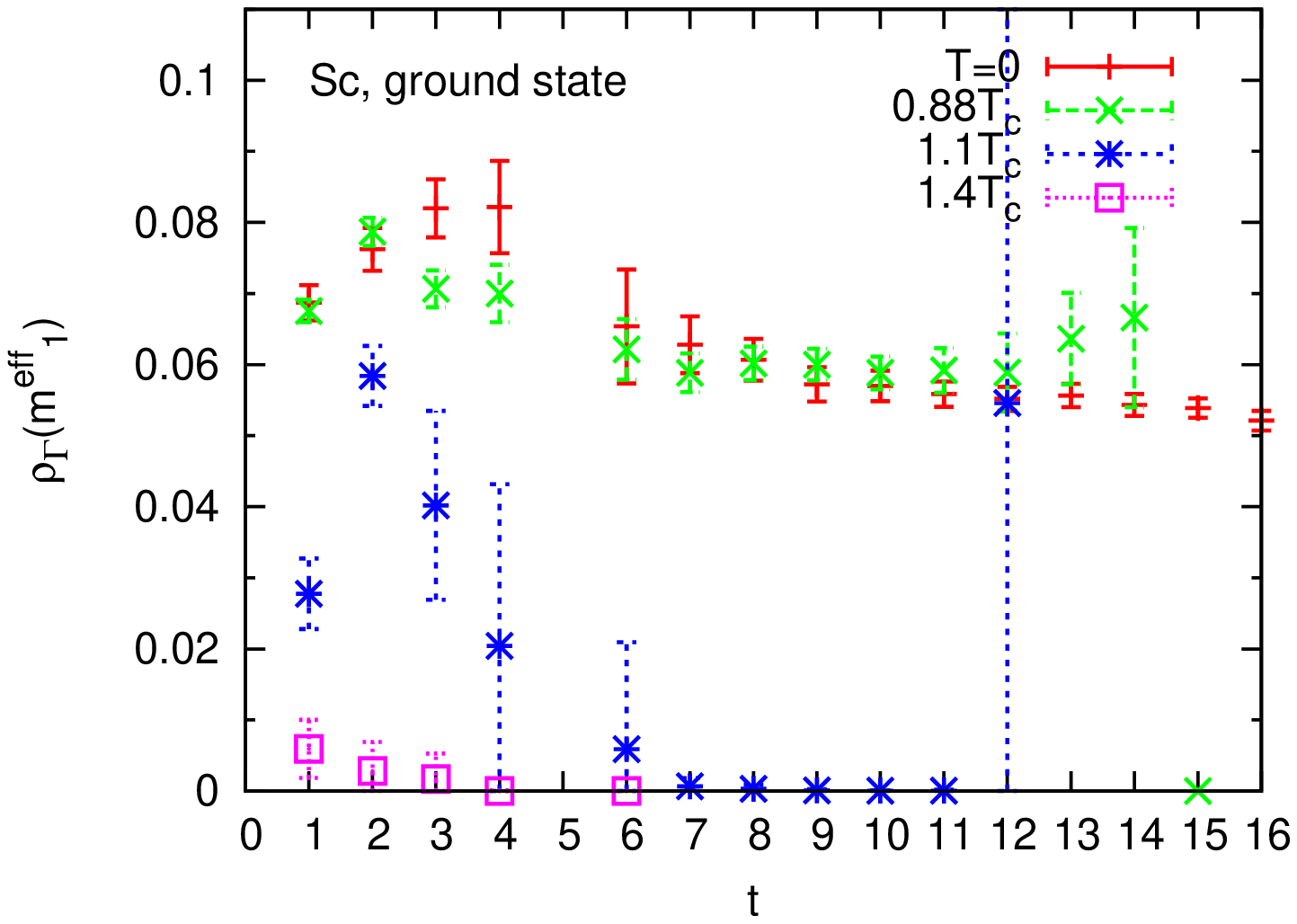}
  \caption{$m^{\mathrm{eff}}_1(t,t_0)$ and $\rho_{\Gamma}(m^{\mathrm{eff}}_1(t,t_0))$ for the Sc channel.} 
  \label{temp_mass_Sc}
 \end{center}
\end{figure}

\subsection{\label{sec:finite_temp}Finite temperature}

Finally, we study the charmonia spectral functions at finite temperature.
We calculate $m_k$ and $\rho_{\Gamma}(m_k)$ with the variational method using the same smearing functions with $n=7$ and $t_0=5$.

In Figs.~\ref{temp_mass_Ps} and \ref{temp_mass_Ve}, we show the results of $m^{\mathrm{eff}}_1(t,t_0)$ and $\rho_{\Gamma}(m^{\mathrm{eff}}_1(t,t_0))$ for the Ps and Ve channels at $T=0$, $0.88T_c$, $1.1T_c$, and $1.4T_c$. 
We see no clear temperature dependence in $m^{\mathrm{eff}}_1$ (upper panels) up to $1.4T_c$. 
On the other hand, $\rho_{\Gamma}(m^{\mathrm{eff}}_1)$ (lower panels) show a slight shift between the temperatures below and above $T_c$. 
However, the temperature dependence is not a drastic one as expected when a particle dissociates.
Thus, these results suggest that both $\eta_c$ and $J/\psi$ survive up to $1.4T_c$.
Because the effective masses and effective spectral functions are not asymptotic up to the largest $t$ available, we do not attempt to fit a plateau. 

The results for the  first excited state Ps channel are shown in Fig.~\ref{temp_mass_Ps1}.
We also found that results for the Ve channel look similar.
We note that both the effective mass and effective spectral function show strong $T$-dependence above $T_c$.
The results for P-waves are also similar to those for the S-wave first excited states.
Even for the ground states, as shown in Fig.~\ref{temp_mass_Sc} for the Sc channel, we find the strong $T$-dependence above $T_c$. 
This may suggest appearances of additional poles.
However, since we cannot extract asymptotic signals from the present data \footnote{We observe that these quantities also show the strong dependence on $n$.},  it is difficult to draw a definite conclusion on the fate of corresponding charmonia above $T_c$.  

\section{\label{sec:conclusions}Conclusions}

We introduced a method to calculate meson spectral functions with the variational method.
We first confirmed by a test in the free quark case that the method reproduces the analytic solutions well for several low-lying states when the number of trial operators, $n$, is sufficiently large.
By introducing more trial operators, we can systematically improve the signal.
On the other hand, a judicious choice of the trial operators is needed to obtain an asymptotic signal within the range of available $t$, in particular, for highly excited states in P-waves.
This imposes a severe limitation on the applicability of the method at high temperatures.
A good feature of the method is, however, that we can judge to which extent the results are regarded as asymptotic by examining the $t$- and $n$-dependences of effective masses and effective spectral functions.

We then adopted the variational method to calculate charmonium spectral functions in quenched QCD.
Comparing the results of the variational method with those of the conventional MEM at zero temperature, we found that the location and the area of the ground state peak by MEM are well reproduced by the variational method.
For the first excited states, we find that the variational method leads to spectra much closer to the experimental ones.
We note that the results of MEM approximately corresponds to those of the variational method in the limit of small $n$.

We also studied the temperature dependence of spectral functions.
We found that the effective masses for the S-wave ground states show no $T$-dependence up to $1.4 T_c$.
Corresponding effective spectral functions show a slight shift between below and above $T_c$.
The absence of a drastic $T$-dependence suggests that $\eta_c$ and $J/\psi$ do not dissociate up to $1.4 T_c$.
However, the limitation in the range of $t$ poses a severe constraint to extract asymptotic signals.

In systems with finite volume, continuum spectra in the infinite-volume limit must break up into discrete spectra.
In the case of free quarks, we have confirmed the appearance of expected discrete spectra in the spectral function for meson operators.
In QCD at $T=0$, we do not expect such additional discrete spectra around the ground state because they appear only above the two-particle threshold in the same channel.
Accordingly, we did not observe them around the states we studied.
When a meson dissociates at high temperature, we expect that the corresponding discrete spectrum changes into a broad continuous peak in the spectral function at that temperature in the infinite-volume limit. 
On finite lattices, we thus expect the appearance of additional discrete spectra around the original spectrum. 
We have observed strong $T$-dependence of the signal at intermediate distances in several channels above $T_c$, which may be suggesting the appearance of these additional spectra in these channels.
To draw a definite conclusion, however, we need asymptotic signals for these spectra.
When such asymptotic signals become available, it is important to check the volume dependence and the overlap with the meson operators used.
More work is needed, in particular, in optimizing the trial operators, to discuss the fate of charmonia above $T_c$.

\begin{acknowledgments}
We thank other members of the WHOT-QCD Collaboration for valuable discussions.
This work is in part supported 
by Grants-in-Aid of the Japanese Ministry of Education, Culture, Sports, Science and Technology, 
(Nos.~20340047, 21340049, 22740168, 22840020)
and by the Grant-in-Aid for Scientific
Research on Innovative Areas (No. 2004: 20105001, 20105003).
HO is supported by the Japan Society for the Promotion of Science for Young Scientists.
The simulations have been performed on a supercomputer NEC SX-8 at the Research Center for Nuclear Physics (RCNP) at Osaka University.
\end{acknowledgments}

\appendix
\section{\label{sec:analytic_SPF}Analytic solution of meson spectral functions for free Wilson quarks}

\begin{table}[t]
\caption{$a$, $b_j$ and $c$ defined by (\ref{eq:traces}) for Ps, Ve, Sc, Av channels, where $i$ is the spatial direction of $\Gamma$ for Ve and Av channels.}
\label{traces}
\begin{ruledtabular}
\begin{tabular}{c|cccc}
   & $\Gamma$                 & $a$ & $b_{j}$        & $c$  \\ \hline
Ps & $\gamma_5$               & $1$ & $0$            & $0$  \\
Ve & $\gamma_i$               & $1$ & $\delta_{ij}$  & $0$  \\
Sc & $\textrm{\boldmath $1$}$ & $0$ & $1$            & $1$ \\
Av & $\gamma_5\gamma_i$       & $0$ & $1-\delta_{ij}$ & $1$ \\
\end{tabular}
\end{ruledtabular}
\end{table}

The free quark propagator for the Wilson quarks in the momentum space is given by
\begin{equation}\label{eq:quark_prop}
S(p) = \frac{-i\gamma_{4} \sin p_{4} - i\mathcal{P}(\vec{p}) + 1-\cos p_{4} + \mathcal{M}(\vec{p})}{\sin^2 p_{4} + \mathcal{P}^2(\vec{p}) + [1-\cos p_{4} + \mathcal{M}(\vec{p})]^2}
\end{equation}
with
\begin{eqnarray}
\mathcal{P}(\vec{p}) &\equiv& \frac{1}{\xi} \sum^{3}_{j=1}\gamma_{j} \sin p_{j}, \nonumber \\
\mathcal{M}(\vec{p}) &\equiv& \frac{1}{\xi} \left[r\sum^{3}_{j=1}(1-\cos p_{j}) + \hat{m} \right]
\end{eqnarray}
where $\xi=a_s/a_t$ is the lattice anisotropy \cite{free_prop}.
With the antiperiodic boundary condition in the temporal direction,
the Fourier transform of the quark propagator in $p_4$ reads
\begin{eqnarray}\label{eq:quark_prop2}
S(\vec{p},t) &=& \gamma_{4} S_{4}(\vec{p}) \cosh[E(\vec{p})(t-N_t/2)] \nonumber \\
 &+& \left[i\sum^{3}_{j=1}\gamma_{j} S_{j}(\vec{p}) + \textrm{\boldmath $1$}S_{u}(\vec{p}) \right] \sinh[E(\vec{p})(t-N_t/2)] \nonumber \\
 &-& \frac{\textrm{\boldmath $1$}\delta_{t0}}{2(1+\mathcal{M}(\vec{p}))},
\end{eqnarray}
where
\begin{eqnarray}
S_{4}(\vec{p}) &\equiv& \frac{\sinh E(\vec{p})}{2(1+\mathcal{M}(\vec{p}))\sinh E(\vec{p}) \cosh[E(\vec{p}) N_t/2]}, \nonumber \\
S_{j}(\vec{p}) &\equiv& \frac{\sin p_{j}}{2(1+\mathcal{M}(\vec{p}))\sinh E(\vec{p}) \cosh[E(\vec{p}) N_t/2]}, \nonumber \\
S_{u}(\vec{p}) &\equiv& - \frac{1-\cosh E(\vec{p}) + \mathcal{M}(\vec{p})}{2(1+\mathcal{M}(\vec{p}))\sinh E(\vec{p}) \cosh[E(\vec{p}) N_t/2]}. \nonumber \\
&&
\end{eqnarray}
Here, $E(\vec{p})$ is the location of the pole of (\ref{eq:quark_prop}):
\begin{equation}
\cosh E(\vec{p}) = 1 + \frac{\mathcal{P}^{2}(\vec{p})+\mathcal{M}^{2}(\vec{p})}{2(1+\mathcal{M}(\vec{p}))}.
\end{equation}

Thus, the meson correlation functions (\ref{eq:meson_corr}) in the free quark case are given by
\begin{equation}
C_{\Gamma}(t) = \frac{N_{c}}{N^{3}_{s}} \sum_{\vec{p}} \mathrm{tr}[\Gamma S(\vec{p},t)\Gamma^{\dag} \gamma_{5} S^{\dag}(\vec{p},t) \gamma_{5}]
\end{equation}
for flavor nonsinglet channels.
Substituting (\ref{eq:quark_prop2}), we obtain
\begin{eqnarray}
C_{\Gamma}(t) &=& \frac{N_c}{N^{3}_{s}} \sum_{\vec{p}}\frac{1}{(1+\mathcal{M}(\vec{p}))^{2}\cosh^{2}[E(\vec{p})N_t/2]} \nonumber \\
 &\times& \left[\left\{a - \sum^{3}_{j=1}b_{j}\frac{\sin^{2}p_{j}}{\sinh^{2}E(\vec{p})}\right\}\cosh[2E(\vec{p})(t-N_t/2)] \right. \nonumber \\
 &-& \left. \left\{c - \sum^{3}_{j=1}b_{j}\frac{\sin^{2}p_{j}}{\sinh^{2}E(\vec{p})}\right\} \right]
 \label{eq:A7}
\end{eqnarray}
for $t>0$, where $a$, $b_j$ and $c$ are calculated by traces of $\gamma$ matrices as
\begin{eqnarray}\label{eq:traces}
a &=& \frac{1}{8}\mathrm{tr}\{\Gamma \Gamma^{\dag} \} - \mathrm{tr}\{\Gamma \gamma_4 \Gamma^{\dag} \gamma_4 \}], \nonumber \\
\nonumber \\
b_j &=& \frac{1}{8}[\mathrm{tr}\{\Gamma \Gamma^{\dag} \} + \mathrm{tr}\{\Gamma \gamma_j \Gamma^{\dag} \gamma_j \}], \nonumber \\
c &=& \frac{1}{8}[\mathrm{tr}\{\Gamma \Gamma^{\dag} \} + \mathrm{tr}\{\Gamma \gamma_4 \Gamma^{\dag} \gamma_4 \}].
\end{eqnarray}
Concrete values of them are summarized in TABLE \ref{traces}.

The last term in (\ref{eq:A7}) is removed by the midpoint subtraction procedure (\ref{eq:midpoint}).
For the Ps channel, the last term is absent because $b_j=c=0$ as listed in Table~\ref{traces}.
On the other hand, for the P-wave states (the Sc and Av channels), the last term is numerically large. 

Adopting the midpoint subtraction procedure, the analytic solution for the meson spectral function in the free Wilson quark case is now given by
\begin{eqnarray}
\tilde\rho_{\Gamma}(\omega) &=& 
\frac{N_c}{N^{3}_{s}} \sum_{\vec{p}} 
\frac{\sinh[E(\vec{p}) N_t]}{(1+\mathcal{M}(\vec{p}))^{2}\cosh^{2}[E(\vec{p}) N_t/2]} 
\nonumber \\
 &\times& \left\{a - \sum^{3}_{j=1}b_{j}\frac{\sin^{2}p_{j}}{\sinh^{2} E(\vec{p})}\right\} 
 \delta(\omega-2E(\vec{p})). 
 \nonumber \\
 &&
\end{eqnarray}

\begin{table}[tbh]
\caption{$t_0$ for the $20^3 \times 128$ lattice}
\label{t0:20128}
\begin{center}
\begin{tabular}{cc|cccccccc} \hline
   && $n=3$ & $4$ && $5$ && $6$ && $7$ \\ \hline
Ps && 62 & 62 && 62 && 60 && 53 \\
Ve && 62 & 62 && 62 && 56 && 52 \\
Sc && 62 & 62 && 57 && 25 && 13 \\
Av && 62 & 62 && 57 && 25 && 13 \\ \hline
\end{tabular}
\end{center}
\end{table}

\begin{figure}[tbh]
 \begin{center}
  \includegraphics[width=55mm, angle=-90]{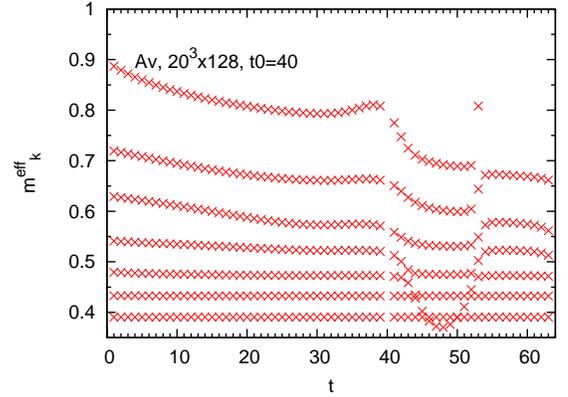}
  \caption{Effective mass for the Av channel on the $20^3 \times 128$ lattice with $t_0=40$. All the seven levels for $n=7$ are shown as functions of $t$.} 
  \label{level_crossing}
 \end{center}
\end{figure}

\section{\label{free_SPF}Effective meson masses and meson spectral functions in the free quark case}

As discussed in  Sec.~\ref{sec:free}, we choose $t_0$ as large as possible in the region where signals of $m^{\mathrm{eff}}_k$ and $\rho(m^{\mathrm{eff}}_k)$ up to the second excited states ($k=1$, 2, 3) are both stable for all $t$ and $n$ up to $n=7$.
We use $t=N_t/2-1$ to extract asymptotic values of $m_k$ and $\rho_{\Gamma}(m_k)$.
Our choices of $t_0$ for the $20^3 \times 128$ lattice are summarized in Table~\ref{t0:20128}.
On the $20^3 \times 32$ lattice, we adopt $t_0=14$ for all channels and all values of $n$.

Results of the effective mass $m^{\mathrm{eff}}_k$ and the effective spectral function $\rho_{\Gamma}(m^{\mathrm{eff}}_k)$ as functions of $t$ and $n$ are given in Figs.~\ref{t_dep_Ps:20128} and \ref{t_dep_Sc:20128} for the Ps and Sc channels.
Results for the Ve and Av channels are similar to those for the Ps and Sc channels, respectively.

When we adopt $t_0$ larger than the value given in Table~\ref{t0:20128}, we encounter unstable signals at several intermediate values of $t$. 
An example is shown in Fig.~\ref{level_crossing}.
We find that a strange level appears at intermediate values of $t$.
As we vary $t$, the strange level crosses the ordinary levels which have milder dependences on $t$. 
We find that the strange levels are suppressed when we adopt a sufficiently small $t_0$, or limit ourselves to smaller values of $n$.
These strange levels disturb the naming of low-lying states and the reliability of the signals. 
To avoid such levels for all $n$ up to $n=7$, we adopt the values of $t_0$ listed in Table~\ref{t0:20128}.

\end{document}